\documentclass{aa}

\usepackage{graphicx}
\usepackage{rotating}
\usepackage{txfonts}
\usepackage{natbib}

\begin{document}

\title{The VMC survey -- LIII. Data release \#7}

\subtitle{Complete survey data and data from additional programmes}

\author{M.-R. L. Cioni\inst{1}
	\and N. J. G. Cross\inst{2}
	\and V. Ripepi\inst{3}
	\and L. Girardi\inst{4}	
	\and R. de Grijs\inst{5,6,7}	
	\and G. Clementini\inst{8}
	\and J. Th. van Loon\inst{9}
	\and M. A. T. Groenewegen\inst{10}		
	\and V. D. Ivanov\inst{11}
	\and J. M. Oliveira\inst{9}		
	\and M. Marconi\inst{3}	
	\and K. Bekki\inst{12}
	\and R. P. Blake\inst{2}
	\and M. J. Holliman\inst{2}
	\and J. Irwin\inst{13}
	\and M. J. Irwin\inst{13}
	\and R. G. Mann\inst{2}
	\and M. Read\inst{2}
	\and E. Sutorius\inst{2}
	\and J. E. M. Craig\inst{9}
	\and F. Cusano\inst{8}
	\and F. Dresbach\inst{9}
	\and C. Maitra\inst{14}	
	\and R. Molinaro\inst{3}
	\and F. Niederhofer\inst{1}
	\and A. O. Omkumar\inst{1,15}
	\and C. M. Pennock\inst{2}
	\and T. Sicignano\inst{11,17,3,18}	
	\and S. Subramanian\inst{16}
	\and B. L. Tatton\inst{9}
	\and S. Vijayasree\inst{1,15}
	\and S. Zaggia\inst{8}
	}	
	
\institute{Leibniz Institut f\"{u}r Astrophysik Potsdam, An der Sternwarte 16, D-14482 Potsdam, Germany\\
		\email{mcioni@aip.de}
	\and Institute for Astronomy, Royal Observatory,  Blackford Hill, Edinburgh EH9 3HJ,  United Kingdom
	\and INAF -- Osservatorio Astronomico di Capodimonte, salita Moiariello 16, I-80131 Napoli, Italy
	\and INAF -- Osservatorio Astronomico di Padova, Vicolo dell'Osservatorio 5, I-35122 Padova, Italy
	\and School of Mathematical and Physical Sciences, Macquarie University, Balaclava Road, Sydney, NSW 2109, Australia
	\and Astrophysics and Space Technologies Research Centre, Macquarie University, Balaclava Road, Sydney, NSW 2109, Australia
	\and International Space Science Institute--Beijing, 1 Nanertiao, Zhongguancun, Hai Dian District, Beijing 100190, China
	\and 	INAF -- Osservatorio di Astrofisica e Scienza dello Spazio, Via Gobetti 93/3, I-40129 Bologna, Italy
	\and Lennard-Jones Laboratories, Keele University, Keele ST5 5BG, United Kingdom
	\and Koninklijke Sterrenwacht van Belgi\"{e}, Ringlaan 3, B-1180 Brussels, Belgium
	\and European Southern Observatory, Karl-Schwarzschild-Strasse 2, D-85748 Garching bei M\"{u}nchen, Germany
	\and ICRAR, M468, The University of Western Australia, 35 Stirling Hwy, Crawley, Western Australia 6009, Australia
	\and Institute of Astronomy, University of Cambridge, Madingley Road, Cambridge CB3 0HA, United Kingdom
	\and Max-Planck-Institut f\"{u} Extraterrestrische Physik, Giessenbachstrasse 1, D-85748 Garching bei M\"{u}nchen, Germany
	\and Institut f\"{u}r Physik und Astronomie, Universit\"{a}t Potsdam, Haus 28, Karl-Liebknecht-Str. 24/25, D-14476 Potsdam, Germany
	\and Indian Institute for Astrophysics, II Block Koramangala, Bengaluru 560 034, India
	\and Scuola Superiore Meridionale, Largo S.\,Marcellino 10, I-80138 Napoli, Italy
	\and Instituto Nazionale di Fisica Nucleare (INFN), Sezione di Napoli, Via Cinthia 21, I-80126 Napoli, Italy
	}	
	
\date{Received Februrary 25, 2025; accepted April 25, 2025}

\abstract{The near-infrared $YJK_\mathrm{s}$ Visual and Infrared Survey Telescope for Astronomy (VISTA) survey of the Magellanic Clouds (VMC) is complete and there are also data from additional programmes enhancing its quality over the original footprints.}{This work presents the final data release of the VMC survey, which includes additional observations and provides an overview of the scientific results. The overall data quality has been revised and reprocessed standard data products, that already appeared in previous data releases, are made available together with new data products. These include individual stellar proper motions, reddening towards red clump stars and source classifications. Several data products, such as the parameters of some variable stars and of background galaxies, from the VMC publications are associated to a data release for the first time.}{The data are processed using the VISTA Data Flow System and additional products, e.g. catalogues with point-spread function photometry or tables with stellar proper motions, are obtained with software developed by the survey team.}{This release supersedes all previous data releases of the VMC survey for the combined (deepstacked) data products, whilst providing additional (complementary) images and catalogues of single observations per filter. Overall, it includes about 64 million detections, split nearly evenly between sources with stellar or galaxy profiles.}{The VMC survey provides a homogeneous data set resulting from deep and multi-epoch $YJK_\mathrm{s}$-band imaging observations of the Large and Small Clouds, the Bridge and two fields in the Stream. The VMC data represent a valuable counterpart for sources detected at other wavelengths for both stars and background galaxies.}

\keywords{Surveys -- Magellanic Clouds -- Infrared: stars}

\maketitle

\section{Introduction}
Near-infrared observations are particularly suited to capture red sources like evolved giant stars with peaks of energy distributions at $\sim$1 $\mu$m, sources behind dust or red-shifted distant galaxies. These are important tracers of structures within the Milky Way, our nearest galaxies and of the comic web. The Clouds system (see \citealp{dennefeld2020} for a history about the naming of the system), which includes two interacting star-forming galaxies at $\sim$50 kpc (e.g., \citealp{degrijs2014}, \citealp{degrijs2015}) and their tidal features (e.g. the Bridge and Stream), can be studied both globally and in detail. 
Stars in the Large Magellanic Cloud (LMC) and the Small Magellanic Cloud (SMC) are fundamental benchmarks for stellar properties and stellar evolution because of their relatively low metallicity. These galaxies are also crucial calibrators of the distance scale through the Leavitt's Law, the relation between the brightness and the pulsation period of Cepheid stars (e,g, \citealp{madore2024}). 

The Visual and Infrared Survey Telescope for Astronomy (VISTA; \citealp{sutherland2015}) imaged the southern sky for about thirteen years and the observation of the Clouds was among the first set of targets endorsed by the European Southern Observatory (ESO). The VISTA survey of the Magellanic Clouds (VMC), described in \cite{cioni2011}, is the most sensitive and high spatial resolution near-infrared imaging survey to date of the Clouds system. This corresponds to detecting (on average) sources with $K_\mathrm{s}$=19.3 mag with an uncertainty <0.1 mag at a resolution of <1 arcsec. Data were collected from 2009 to 2018 during the equivalent of 2000 hours or 250 nights, mapping the system with >100 VISTA tiles, each tile is $\sim$1.5 deg$^2$ in size (at the LMC distance, 1 deg corresponds to $\sim$1 kpc). There have been six public data releases and the VMC team has produced over 60 articles in major astronomical journals on a wide range of scientific topics. Major results include recovering a spatially resolved star formation history (SFH) and mapping the structure of the galaxies in three dimensions (3D) using multiple tracers, as well as separating the kinematics (from proper motion) of young and old stellar populations in their inner and outer regions. Twelve additional imaging programmes, lead by the VMC team (except for two) and complementing the VMC survey, were completed before the VISTA camera was removed from the telescope in early 2023. These programmes add multi-epoch observations to the VMC foot-print. 

This work presents the final VMC data release, which includes the reprocessing of the entire set of VMC data combined with data from the additional programmes. This combination allows to reach fainter sources and add new epochs to measure proper motions or study variable stars better than by using VMC data alone. It also features an overall re-evaluation of the imaging quality by applying stringent criteria to produce deep (combined) data products, a table of proper motions of individual stellar sources (for the first time) and tables with parameters that have already appeared in journal publications or are about to do so, but that have not yet been linked to the VMC data products through a public data release. This is the only VMC paper presenting an overview of the data release because previous studies have focused on specific science goals. Details about the observations are given in Sect.~\ref{observations} while the data processing is described in Sect.~\ref{process} and the data products are presented in Sect.~\ref{data}.  A summary of the VMC scientific results is given in Sect.~\ref{summary} whereas Sect.~\ref{conclusions} concludes the final data release overview.

\section{Observations}
\label{observations}

\subsection{The VMC programme}
\label{vmc}

VMC observations refer to the ESO programme identification 179.B$-$2003. They started in October 16, 2009 and were completed in October 16, 2018 using the 4.1 m VISTA telescope \citep{sutherland2015} located at ESO's Cerro Paranal Observatory in Chile and the VISTA infrared camera (VIRCAM; \citealp{dalton2006}, \citealp{emerson2006}). VIRCAM is equipped with 16 detectors of 2048$\times$2048 pixels each arranged in a 4$\times$4 pattern with a physical separation of 42.5\% in the Y direction and 90\% in the X direction. An image with this configuration is named a {\it pawprint} and a mosaic of six {\it pawprints} is necessary to produce a contiguous image (tile) of the sky. A tile covers an area of about 1.77 deg$^2$ where the central $1.5$ deg$^2$ is observed at least twice due to the detector overlaps and two stripes at the edges only once. In the VMC mosaic, tiles overlap such that these underexposed sides are also observed twice while the overlap for the other sides is about 30 arcsec. Detector and tile overlaps may vary in size as a result of the automatic allocation of guiding/reference stars. SMC, Bridge and Stream tiles follow the default orientation, where the Y axis points to the North and the X axis to the West, while LMC tiles are oriented at a position angle of +90 deg; see \cite{cioni2011} for the construction of the VMC mosaic and App.\,\ref{maps} for the maps. Note that soon after the first observational season it was decided to remove tile LMC 11\_6 on the North and introduce tile LMC 7\_1 on the West (cfg. Fig.\,A.1 in \citealp{cioni2011} with Fig.\,\ref{lmcmap} in this study).

VMC observations cover a total sky area of about 171.5 deg$^2$ (110 tiles) of which 104.8 deg$^2$ on the LMC (68 tiles; Fig.\,\ref{lmcmap}), 42.1 deg$^2$ on the SMC (27 tiles; Fig.\,\ref{smcmap}), 21.1 deg$^2$ on the Bridge (13 tiles; Fig.\,\ref{bridgemap})  and 3.5 deg$^2$ on the Stream (2 tiles). Images were acquired in the $Y$ ($\lambda_\mathrm{center}$=1.02 $\mu$m), $J$ ($\lambda_\mathrm{center}$=1.25 $\mu$m) and $K_\mathrm{s}$ ($\lambda_\mathrm{center}$=2.15 $\mu$m) broad band filters with at least two epochs in the $Y$ and $J$ bands (providing 800 s exposure time per pixel each)  and 11 in the $K_\mathrm{s}$ band with 750 s exposure time per pixel each. These are {\it deep} epochs and they constitute the majority of the observations. In addition, there are two {\it shallow} epochs for each tile and in each band, with about half of the exposure time of the deep ones, taken to extend the non-linear dynamic range of the observations. Pairs of shallow epochs ($YJ$, $JK_\mathrm{s}$ and $YK_\mathrm{s}$) were observed consecutively with the purpose to reduce the impact of variability on colours. Deep epochs in the $Y$ and $J$ bands had no time restrictions; they were sometimes observed in the same band during the same night. On the other hand, deep epochs in the $K_\mathrm{s}$ band had a minimum time separation of 1, 3, 5, 7 and 17 days for each subsequent epoch. This cadence was specifically designed to capture both the short and long-time scale variability of pulsating stars like RR Lyrae stars and Cepheids, respectively. Further details about the observing strategy are given in \cite{cioni2011}.

\subsection{Additional programmes}
\label{addition}

Several observations of the VMC tiles were obtained outside the nominal time of the survey and as part of open-time programmes to address specific science questions, complement the survey and enhance its legacy value. Their ESO programme identifications are: 099.C$-$0773, 099.D$-$0194, 0100.C$-$0248, 0103.B$-$0783, 0103.D$-$0161, 105.2042, 106.2107, 108.222A, 108.223E, 109.230A, 109.231H, and 110.259F; they are described in detail below. The Clouds were also observed during the VISTA commissioning and as part of a target-of-opportunity programme on compact binaries (095.D$-$0771). These observations are however not included here because of their significantly different observing strategies compared with the VMC programme.

\subsubsection{Probing the variability of young stars}
\label{zivkov}

Programmes 099.C$-$0773 and 0100.C$-$0248 were designed to almost double the number of epochs in both $J$ and $K_\mathrm{s}$ bands for two VMC tiles, SMC 5\_4 (Fig.\,\ref{smcmap}) and LMC 7\_5 (Fig.\,\ref{lmcmap}), with the goal of studying the variability of young stars. These two tiles were chosen to maximise the coverage of areas with a strong recent star formation activity (\citealp{rubele2015,harris2009}) and because they contain more than fifty young stellar clusters (\citealp{glatt2010,chiosi2006}) with ages below 10 Myr. Photometric variability of pre-main sequence stars is closely connected to early stellar evolution. For instance, episodic changes in circum-stellar mass accretion rates lead to eruptive highly variable young stellar objects (YSOs), while structural asymmetries in the inner disc cause semi-periodic brightness -- this is because these structures may come and go, the phases would change; also the structures can evolve, so the amplitudes may also change. The additional $K_\mathrm{s}$ epochs obtained with these programmes cover the timescale of days up to one month and provide contemporaneous observations in the $J$ band. A comparison between the LMC and SMC allows also to study any variability dependence on metallicity. 

The 0100.C$-$0248 programme added 12 epochs between January 17, 2018 and February 6, 2018 to the VMC epochs for the LMC 7\_5 in both the $J$ and $K_\mathrm{s}$ bands, resulting in a combined time baseline of 6$-$7 years. The exposure time of the additional observations is longer in $K_\mathrm{s}$ (480 s versus 175 s and 375 s) and shorter in $J$ (90s versus 200 s and 400 s) than the one in the VMC survey for shallow and deep epochs. This is achieved by varying the number of detector integration time (NDIT). The jittern pattern ({\it jitter3u}) also differs from that adopted in the VMC survey ({\it jitter5n}). The other  parameters: DIT (5 s), number of exposures (1) and pawprints (6), micro-stepping (1) and tile pattern ({\it Tile6zz}) are identical with those in the VMC survey. The exposure time per pawprint is calculated as DIT$\times$NDIT$\times$(number of jitters). Details about the observing strategy for programme 0100-C248 are given in \cite{zivkov2020}.

The 099.C$-$0773 programme added 14 epochs between July 6, 2017 and August 9, 2017 to the VMC epochs for the tile SMC 5\_4 and adopted the same parameters as for programme 0100-C$-$0248. The combined time baseline is of similar length in both programmes. However, the tile pattern ({\it Tile6zz}) differs from the one adopted by the VMC survey ({\it Tile6n}) for observing the SMC.

\subsubsection{Filling in the gaps in the LMC and SMC observations}
\label{gapsection}

Programmes 099.D$-$0194, 0103.D$-$0161, 105.2042, and 109.230A were designed to fill in a gap left by VMC observations in the footprint of the SMC, whereas programme 108.223E was similarly aimed at filling in a gap in the LMC. A lack of VISTA observations in the gap regions influences the study of substructures within the galaxies, by introducing artificial discontinuities (see \citealp{elyoussoufi2019}), the derivation of surface density profiles and radial profiles/gradients as well as the spatially resolved kinematics of stellar populations. In \cite{sun2018} the SMC-gap region was filled with $UBVI$ photometric data from the Magellanic Clouds Photometric Survey (MCPS; \citealp{zaritsky2000,zaritsky2002}) following a scaling and a calibration procedure using the adjacent regions. Likewise, in \cite{miller2022} the LMC-gap region was filled with $UBV$ photometric data from the Survey of MAgellanic Stellar History (SMASH; \citealp{nidever2017, nidever2021}) following the method of \cite{sun2018}. While this worked well for upper main sequence stars, it is not obvious that the different wavelengths and survey sensitivities would satisfactorily describe the spatial distribution of older stellar populations. Furthermore, multi-band data in the inner region of the Clouds are essential to trace interstellar extinction (e.g, \citealp{bell2020,bell2022}).

The SMC gap corresponds to a vertical strip with a length of 1 deg in declination and a width of 0.034 deg in right ascension. It is located in the northern bar region of the galaxy between tiles SMC 5\_3 and 5\_4 (Fig.\,\ref{smcmap}. The gap corresponds to 2.3\% of a VISTA tile in size and contains more than 500 stars with $K_\mathrm{s}$=19.5$-$20 mag. The new observations acquired a new tile centred at the gap (00$:$54$:$58, $-$72$:$00$:$45) following the strategy of the VMC survey. These observations covered the gap and added also extra epochs to the immediate vicinity of it, extending the time coverage of the overlapping area between the new tile and the tiles SMC 5\_3 and SMC 5\_4. All but one (deep) $K_\mathrm{s}$-band epoch were successfully obtained between August 6, 2017 and December 27, 2022.

The origin of the LMC gap is due to a shift in the central coordinates of tile LMC 4\_4 (Fig.\,\ref{lmcmap}) for $YJ$-deep and $YJK_\mathrm{s}$-shallow images compared to the centre of $K_\mathrm{s}$-deep images. Programme 108.223E re-obtained $YJ$-deep and $YJK_\mathrm{s}$-shallow images, following the VMC strategy, at the same location of the existing $K_\mathrm{s}$-deep images effectively covering the gap. These observations required significantly less time than the alternative of re-acquiring all of the $K_\mathrm{s}$-deep images.The LMC gap corresponds to an area of 0.25 deg$^2$ with a length of 1.445 deg in declination and a width of 0.108 deg in right ascension. It is located in the inner region of the galaxy, south of the bar and crossing the south-east spiral arm. The LMC gap corresponds to $\sim$6\% of a VISTA tile in size and contains more than 8\,000 stars with $K_\mathrm{s}$=19.5$-$20 mag. All but one (deep) $Y$-band epoch were successfully obtained between December 16, 2021 and January 30, 2022.

The successful completion of these programmes produced a spatially homogeneous data set to enhance the public scientific value of the VMC survey and its long lasting legacy impact. 

\subsubsection{Improving the measurements of proper motions}
\label{epochs}

Programmes 0103.B$-$0783, 105.2043, 106.2107, 108.222A, 109.231H, and 110.259F were designed to acquire one additional (deep) $K_\mathrm{s}$-band epoch on each VMC tile with the goal to increase the time baseline and improve the measurement of proper motions. The proper motion measured with modern instruments is a powerful tool to characterise kinematic patterns of stellar populations within the Clouds (e.g., \citealp{niederhofer2022}). The combination of the VMC survey epochs with one additional epoch per tile extends the time baseline of uniform observations from about 2 to at least 7 years, depending on when the first and last observation of a given tile were obtained, considerably improving on the capacity to characterise rotation and kinematical substructures within the system. An extra $K_\mathrm{s}$-band epoch is also valuable for long-term variability studies of evolved stars, YSOs, and background Active Galactic Nuclei (AGN). A good quality (deep) $K_\mathrm{s}$-band epoch was obtained for 63 tiles between 2019 August 5 and 2023 January 20 adopting the same parameters as for the VMC survey. Two tiles (LMC 7$\_5$ and SMC 5$\_$4) were not observed in these programmes because sufficient epochs were obtained in the monitoring programmes 099.C$-$0773 and 0100.C$-$0248 (Sect.~\ref{zivkov}).

\section{Data processing}
\label{process}

The processing of the VMC data, from raw images to calibrated images and source catalogues, was performed with the VISTA Data Flow System (VDFS; \citealp{irwin2004}). At the Cambridge Astronomical Survey Unit (CASU\footnote{http://casu.ast.cam.ac.uk}) the individual images, per exposure time and filter, are stacked and combined to deliver astrometrically and photometrically calibrated single pawprints and tiles corresponding to observations at a given epoch, where each epoch corresponds to about a 1 hour long observing sequence. Individual pawprint observations are made of 10 to 75 images, depending on filter and type of epoch (shallow or deep), and there are 6 pawprints making up a tile with 4 to 11 epochs, also depending on filter. Refer to \cite{cioni2011} for details on the number of images, exposure times and repeats. Refer instead to the CASU web pages for details about the specific processing steps: reset, dark, linearity, flat-field and background correction, destriping, jitter stacking, catalogue generation, calibration and tile generation. Note that VISTA detectors are independent, i.e. they have different properties which are corrected at the pixel level during the processing stage. Remaining issues that cannot be resolved/homogenised through the data processing or the survey strategy (for example by allowing for a larger physical overlap to compensate for areas affected by a poor/variable pixel response) are encoded in quality flags. The photometric calibration, which is based on the Two Micron All-Sky Survey (2MASS; \citealp{skrutskie2006}) photometry for stars observed in the VIRCAM pointings, is described in \cite{gonzalez2018} and the precision achieved in the VMC filters ($Y$, $J$, and $K_\mathrm{s}$) is better than 2\%. 

Subsequently, at the Wide Field Astronomy survey Unit (WFAU\footnote{https://ifa.roe.ac.uk/research-areas/wide-field-astronomy-unit}), the pawprints and tiles are further stacked (across epochs) to produce deep pawprints and deep tiles per filter. They are also linked to allow the query of simultaneous data products for sources detected multiple times and at different wavelengths. Individual pass-band detections are merged into multi-colour lists following a procedure\footnote{http://vsa.roe.ac.uk/dboverview.html} based on matching pairs of frames from long ($K_\mathrm{s}$) to short ($Y$) wavelengths, and early to late epochs. The pairing tolerance for the VMC survey is 1 arcsec. This radius is larger than the typical astrometric errors and may introduce some level of spurious matches. Matching objects in the overlap regions of detectors are ranked according to their filter coverage, then their quality error flags and finally their proximity to a detector edge. Note that detections/objects may also be spurious and in this case do not represent astronomical sources. The final band-merged catalogue includes only objects that do not have duplicate measurements, see \cite{hambly2008}. 

The data in DR7, from the VMC survey and the additional programmes, are processed with version 1.5 of the VDFS which includes: an updated photometric calibration, updates to the Galactic extinction coefficients used in generating the photometric zero-points and a fix for a systematic photometric variation across tile catalogues generated prior to January 1, 2017. As a result, magnitude zero-points in single pawprints were updated with changes of the order of 1--2\% compared to previous data processing versions, tile images and source catalogues were regenerated accordingly.

\subsection{Image quality}
\label{imagequality}

Observations for the VMC survey and additional programmes were carried out by ESO staff in service mode, which resulted in a high level of data homogeneity. The mean quality of the combined observations is given in Table \ref{quality}, with standard deviations associated with each parameter. Only tile images of good quality are included in the calculation of the values reported in this table. These are images that meet (within a small tolerance\footnote{https://www.eso.org/sci/observing/phase2/SMGuidelines/\-ConstraintsSet.html}) the requested observational criteria for seeing, sky transparency (THIN or better) and airmass (<1.7). The seeing request, defined as the Full Width at Half Maximum (FWHM) of stellar images, varied with waveband and tile location from 1.0 arcsec to 1.2 arcsec with incremental steps of 0.1 arcsec from the $Y$ and $J$ to the $K_\mathrm{s}$ band. The majority of VMC tiles follow this request, but for 24 tiles covering the densest regions of the galaxies the seeing request was reduced by 0.2 arcsec in each band. The additional programmes follow the same seeing request as for the VMC survey. Some observations, carried out exceptionally down to airmass $\sim$2, for which all other criteria were satisfied, are also included. There were no requirements on the fraction of lunar illumination, since the minimum Moon distance of 60 deg is fulfilled at the location of the VMC tiles. Observations in the $K_\mathrm{s}$ band could also occur up to 30 min into the twilight period because of the reduced sky background in this waveband. Furthermore, we excluded from the calculation of the mean values, presented in Table \ref{quality}, all images of single pawprints that did not result in a completed tile, and tile images for which the corresponding pawprints show zero-point differences $\geq$0.1 mag. 

\begin{table}
	\caption{Combined quality parameters for VMC and additional-programmes observations.}
	\label{quality}
	\small
	\centering
	\begin{tabular}{ccccc}
	\hline\hline
	 Filter & Airmass & FWHM & Ellipticity & Zero Point \\
	  & & ($^{\prime\prime}$) & & (mag) \\
	\hline
	$Y$ & 1.52$\pm$0.07 & 1.04$\pm$0.09& 0.06$\pm$0.01 & 23.48$\pm$0.08 \\
	$J$ & 1.52$\pm$0.07 & 0.97$\pm$0.07& 0.06$\pm$0.01 & 23.71$\pm$0.06 \\
	$K_\mathrm{s}$ & 1.53$\pm$0.07 & 0.92$\pm$0.07 & 0.05$\pm$0.01 & 23.04$\pm$0.03 \\
		 \hline
	\end{tabular}
\end{table}

A complete tile requires six pawprints whereas a complete pawprint requires five images corresponding each to a jitter position.  There are in total 2\,431 good-quality tile images in DR7 which correspond to 14\,586 pawprints and  72\,930 images. Tables \ref{lowquality1} and \ref{lowquality2} list the tile images of low quality within the different components of the VMC survey and for the additional survey programmes, respectively. They provide average quality parameters and a reason for the low quality. Single pawprints of low quality are listed instead in Table \ref{pawprints}. There are in total 626 tiles and 255 single pawprints of low quality in DR7; many of them are suitable for scientific applications.

Tiles usually show spatial variations in depth due to the different properties of the individual detectors, overlapping regions of increased exposure and possible variations of the observing conditions during a given observing sequence, which has a duration of $\sim$1 hour. A dedicated procedure (grouting) is implemented in VDFS to ameliorate these photometric effects. However, the degrading of the VISTA mirrors and the change of coating (from silver to aluminium), that occurred at the beginning of 2011, also impact on the sensitivity of the VISTA images. 
All tiles show a 10--20 mas systematic astrometric pattern due to residual World Coordinate System errors in the pawprints. 
Furthermore, some single jitter images of a stack, making up a pawprint, show that some detectors were swapped, i.e. the detections in a given area appear elsewhere in the field of view. In this case, the resulting tile image will have a reduced sensitivity at the locations of the `missing' detectors. The list of the seven tile images affected by swapped detectors is given in Table \ref{swapped}.

\begin{table*}
	\small
	\setlength{\tabcolsep}{2.5pt}
	\caption{Tile images affected by swapped detectors.}
	\label{swapped}
	\centering
	\begin{tabular}{lcclccccr}
	\hline\hline
	Tile & Date & Filter & Type & FWHM & Ellipticity & Zero Point ($\sigma$) & Airmass & Programme \\
	 & & & & ($^{\prime\prime}$) & & (mag) & & \\
	\hline
	LMC 6\_9 & 2018-09-23 & $K_\mathrm{s}$ & deep & 0.79 & 0.05 & 23.07 (0.01) & 1.671 & 179.B-2003 \\
	LMC 7\_9 & 2018-09-03 & $K_\mathrm{s}$ & deep & 0.95 & 0.06 & 23.07 (0.01) & 1.639 & 179.B-2003 \\
	 & 2018-09-21 & $K_\mathrm{s}$ & deep & 0.88 & 0.05 & 23.07 (0.01) & 1.489 & 179.B-2003 \\
	LMC 9\_8 & 2018-03-10 & $K_\mathrm{s}$ & deep & 0.88 & 0.06 & 22.87 (0.01) & 1.380 & 179.B-2003 \\
	LMC 10\_6 & 2018-03-07 & $K_\mathrm{s}$ & deep & 1.14 & 0.08 & 22.87 (0.01) & 1.406 & 179.B-2003 \\
	SMC 2\_2 & 2019-08-22 & $K_\mathrm{s}$ & deep & 0.88 & 0.07 & 23.03 (0.01) & 1.601 & 0103.B-0783 \\
	SMC 5\_2 & 2019-06-19 & $K_\mathrm{s}$ & deep & 0.92 & 0.09 & 23.02 (0.01) & 1.654 & 0103.B-0783 \\
		 \hline
	\end{tabular}	
\end{table*}

Quality parameters assigned during the post processing at WFAU are listed as quality flags for each detection and sources with only minor quality issues will have ppErrbits\footnote{http://vsa.roe.ac.uk/ppErrBits.html} values <256. Higher values indicate more serious problems, for example that sources lie within the problematic detector \#16 (affected by variable quantum efficiency) or within an underexposed strip of a tile, are close to saturation or correspond to a bright tile detection, but no detection in pawprints.

\section{Data products}
\label{data}

\begin{table}
	\caption{Number of detections.}
	\label{sources}
	\centering
	\begin{tabular}{lrrr}
	\hline\hline
	 Filter(s) & Stars & Galaxies & Noise/Saturated \\
	\hline
	$Y+J+K_\mathrm{s}$ & 21 873 300 & 13 728 100 & 14 000 \\
	$Y+J$ & 6 037 300 & 4 683 000 & 8 900 \\
	$J+K_\mathrm{s}$ & 732 200 & 1 689 100 & 4 300 \\
	$Y+K_\mathrm{s}$ & 347 200 & 568 400 & 1 200 \\
	$Y$ & 2 320 100 & 3 628 500 & 70 100 \\
	$J$ & 1 212 300 & 2 796 800 & 41 700 \\
	$K_\mathrm{s}$ & 738 400 & 3 832 600 & 32 000 \\
	All & 33 260 800 & 30 926 500 &  172 200 \\	
%
		 \hline
	\end{tabular}
\end{table}

\subsection{Standard data products}

The standard data products from the VMC survey and the additional programmes that are released with DR7 consist of images and catalogues processed with the VDFS. For each observation (pointing) there are reduced and calibrated images, in addition to the corresponding pawprints (6 per tile), deep co-added images, confidence maps and catalogues (separately for each filter). The confidence maps reflect the cosmetics of the images and mark regions of poor quality, such as areas of dead pixels, rows and columns not well corrected and the poorly flat-fielded area of detector \#16. For specific examples, see the CASU pages describing the known issues\footnote{http://casu.ast.cam.ac.uk/surveys-projects/vista/technical/known-issues} with the VISTA data. There are also deep co-added tile images (separately for each filter), for both individual tiles and combined, as well as band-merged catalogues, for each tile. The deep-tile products are obtained from combining data of good quality. The observational and data processing parameters are encoded in the FITS headers. Preview images in JPEG format are associated to each FITS image. Celestial coordinates are given at the epoch J2000.0 unless differently specified. Magnitudes are expressed in the VISTA system and are not corrected for extinction; see \cite{gonzalez2018} for conversions to the Vega and AB systems. 

\begin{figure*}
\centering
\includegraphics[width=\hsize]{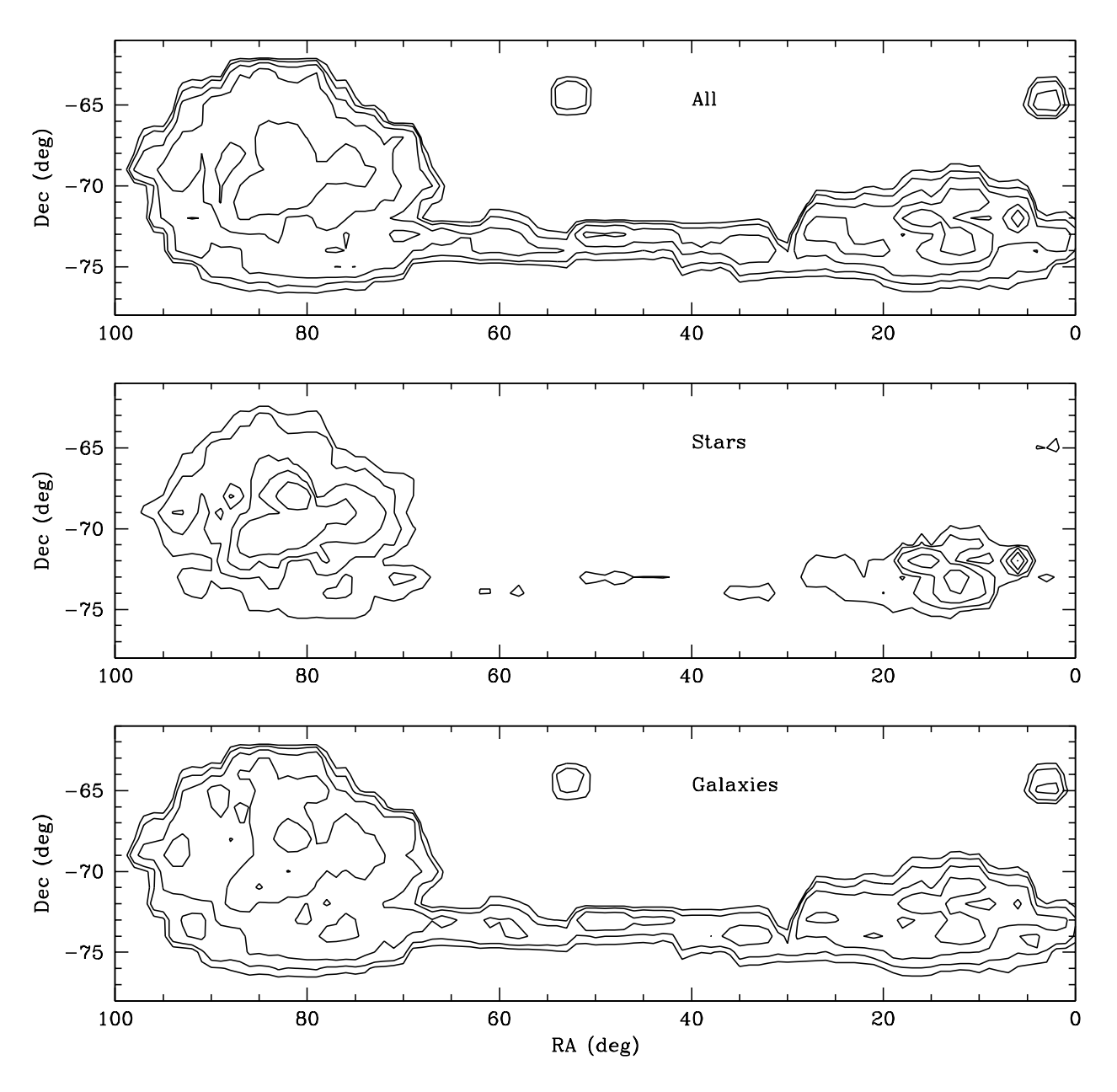}
\caption{Map of VMC detections in $J$ and $K_\mathrm{s}$, without a counterpart in $Y$, (top) with a stellar (middle) and galaxy (bottom) profile. Contours mark number density levels of 500, 1\,000, 2\,000, 4\,000, and 8\,000 sources.}
\label{jkmaps}
\end{figure*}

Table \ref{sources} lists the approximate number of detections in different filters. These are counted by selecting the best detection in the overlap of adjacent tiles since they are processed independently; sources are unique within each tile. There are in total about 64 million entries of which 51.7\% and 48.0\% have a stellar or galaxy profile, respectively, whereas 0.3\% correspond to either noise or saturated sources. At the basis of the morphological classification is the curve-of-growth of the flux of each object and the type of object depends on its sharpness. This process also takes into account the ellipticity and magnitude-dependent information; see \cite{irwin2004} for details. Individual image classifications are combined using Bayesian classification rules as reported in the metadata. Stars prevail over galaxies for detections in three or two filters while the opposite is true for single-band detections. This suggests the presence of populations of sources without obvious counterparts. However, objects detected within dense stellar regions may be mistaken for galaxies if they cannot be disentangled from their neighbours. Elongated objects detected only in one band and in the proximity of bright stars are probably spurious. In general, stars dominate the densest parts of the Clouds whereas galaxies dominate the sparser fields in the Bridge and Stream. This is for example shown in Figure \ref{jkmaps} which illustrates the spatial distribution of sources detected only in $J$ and $K_\mathrm{s}$.
The density of stars at about RA=6 deg and Dec=--72 deg corresponds to the 47 Tucanae (47 Tuc) Galactic globular cluster. The typical features of the SMC are: a North-East and South-West structure within the bar; an elongation to the East towards the Bridge and to the North-West (possibly associated with the Counter Bridge). In the LMC they are: the bar with a Northern overdensity, the 30 Dor star forming region; the Northern spiral arm and substructures in the South towards the Bridge. The number of detections, their sensitivity and spatial distribution depend on selection criteria using source-extraction flags, photometric uncertainties and other catalogue attributes, as well as on specific science applications. Different examples can be found in the published VMC papers. Compared to previous data releases, the present catalogue is more reliable because it contains more observations (from the additional programmes) and is based on stricter data-processing criteria. 

\begin{figure*}
\centering
\includegraphics[width=\hsize]{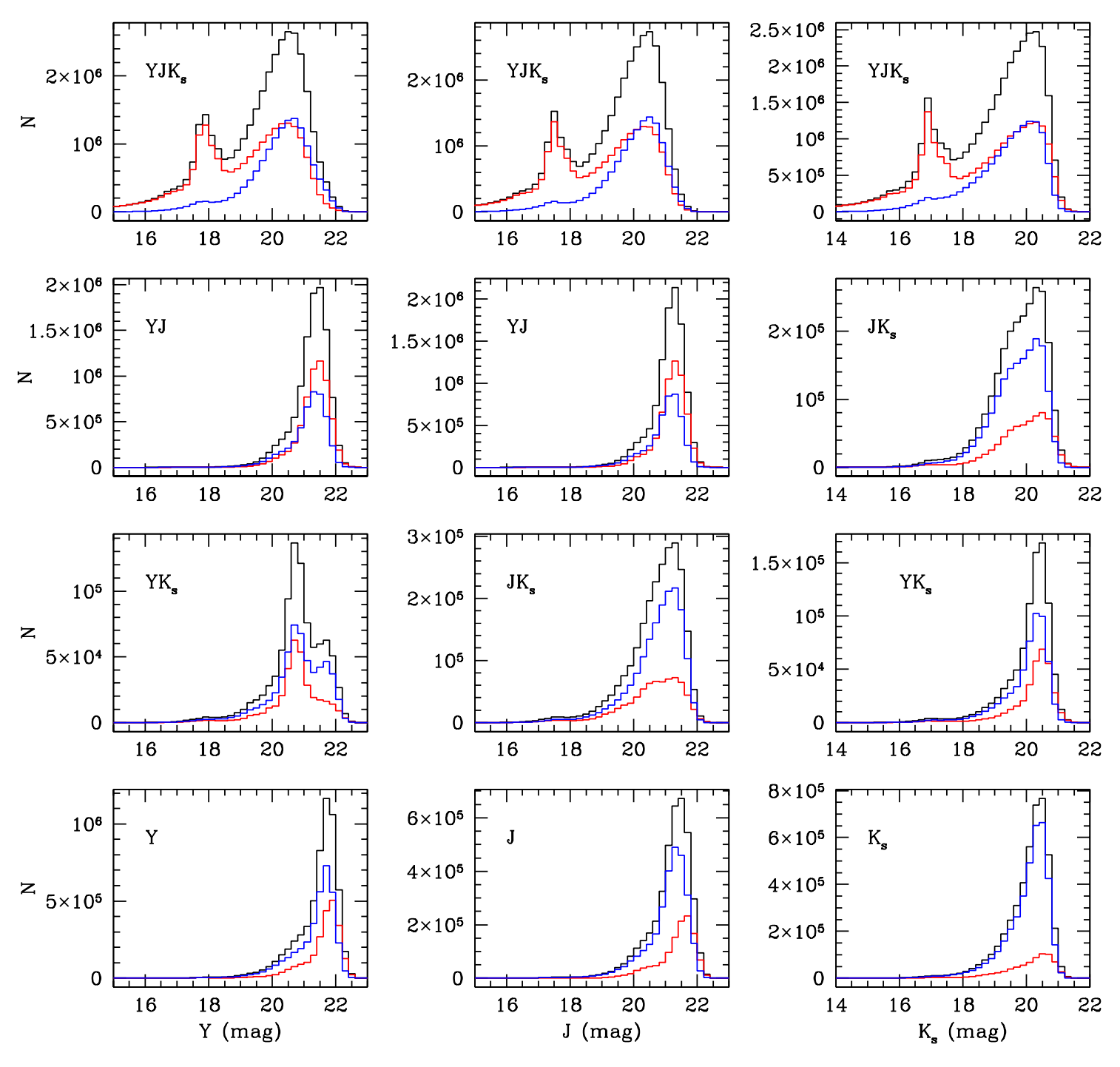}
\caption{Apparent luminosity function in 0.2 mag bins, for the entire catalogue, with detections in multiple (two or three) and single filters. Each column corresponds to a different band and each line corresponds to a different combination of detections, as indicated within each panel. Sources with a stellar profile are shown in red and sources with a galaxy profile are shown in blue whereas their total is shown in black.}
\label{histograms}
\end{figure*}

\begin{figure}
\centering
\includegraphics[width=2.5cm]{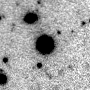}
\includegraphics[width=2.5cm]{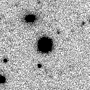}
\includegraphics[width=2.5cm]{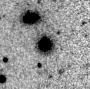}
\includegraphics[width=2.5cm]{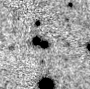}
\includegraphics[width=2.5cm]{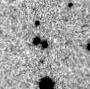}
\includegraphics[width=2.5cm]{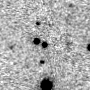}
\includegraphics[width=2.5cm]{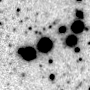}
\includegraphics[width=2.5cm]{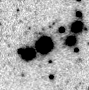}
\includegraphics[width=2.5cm]{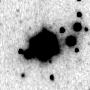}
\includegraphics[width=2.5cm]{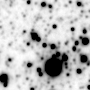}
\includegraphics[width=2.5cm]{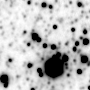}
\includegraphics[width=2.5cm]{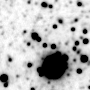}
\includegraphics[width=2.5cm]{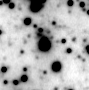}
\includegraphics[width=2.5cm]{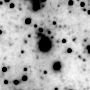}
\includegraphics[width=2.5cm]{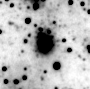}
\includegraphics[width=2.5cm]{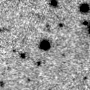}
\includegraphics[width=2.5cm]{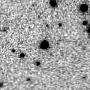}
\includegraphics[width=2.5cm]{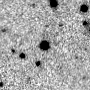}
\includegraphics[width=2.5cm]{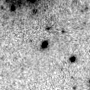}
\includegraphics[width=2.5cm]{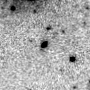}
\includegraphics[width=2.5cm]{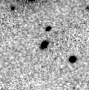}
\caption{Image cut-outs in the $Y$ (left), $J$ (middle) and $K_\mathrm{s}$ filters of different sources in the catalogue. Each image covers an area of 30$\times$30 arcsec$^2$. Each row corresponds to the same central source as follows: (first row) the classical Cepheid OGLE-LMC-CEP-0002 (Table \ref{classical}), (second row) the RR Lyrae star 558384349060 (Table \ref{rrl}), (third row) the LPV star at ($\alpha$, $\delta$)=(5.99372 deg, --73.63190 deg) from Table \ref{agb}, (fourth row) the red clump star at ($\alpha$, $\delta$)=(6.838300 deg, --75.840805 deg) from Table \ref{reddening}, (fifth row) the YSO at ($\alpha$, $\delta$)=(82.5220 deg, --68.6026 deg) from \cite{zivkov2020}, (sixth row) the quasar VMC J001806.53-715554.2 (Table \ref{qso}), and (seventh row) the Scb-type galaxy at ($\alpha$, $\delta$)=(73.2967 deg, --75.69781 deg) and $z\sim$0.2 (Table \ref{bell}).}
\label{stamps}
\end{figure}

Figure \ref{histograms} shows the luminosity functions in different filters and Table \ref{peaks} reports the magnitudes of the highest peaks of the main distributions for each filter combination. Stellar sources detected in three bands show also a secondary peak at 17.9 mag, 17.5 mag and 16.9 mag in the $Y$, $J$ and $K_\mathrm{s}$ band, respectively, at the location of the red clump. Sources detected only in $Y$ and $J$ are probably too faint to show a counterpart at $K_\mathrm{s}$ while many of the sources detected only in $J$ and $K_\mathrm{s}$ are probably galaxies which are too red to show a counterpart in $Y$. Sources detected only in $Y$ and $K_\mathrm{s}$ appear similar to those detected in three bands and the missing $J$ detections might be due to completeness and/or technical reasons. These sources show also a secondary peak at 21.7 mag in the $Y$ band which coincides with the peak of sources that have only a detection in the $Y$ band. Many of the single-band detections in $Y$ are faint stars while in $J$ and especially in $K_\mathrm{s}$ they are probably galaxies with redshifts up to about 3 \citep{bell2020}. Figure \ref{stamps} shows image cut-outs to illustrate typical sources in the catalogue.

\begin{table}
	\small
	\setlength{\tabcolsep}{2.5pt}
	\caption{Magnitude of highest detection peaks per filter combination.}
	\label{peaks}
	\centering
	\begin{tabular}{ccc|ccc|ccc}
	\hline\hline
	\multicolumn{3}{c}{All} & \multicolumn{3}{c}{Stars} & \multicolumn{3}{c}{Galaxies} \\
	 $Y$ & $J$ & $K_\mathrm{s}$  & $Y$ & $J$ & $K_\mathrm{s}$  & $Y$ & $J$ & $K_\mathrm{s}$ \\
	 (mag) & (mag) & (mag) &  (mag) & (mag) & (mag) &  (mag) & (mag) & (mag) \\
	\hline
	  20.5 & 20.5 & 20.3 & 20.5 & 20.3 & 20.3 & 20.7 & 20.5 & 20.1 \\
	  21.5 & 21.3 & & 21.5 & 21.3 & & 21.3 & 21.3 & \\
	   & 21.3 & 20.3 & & 21.3 & 20.5 & & 21.3 & 20.3 \\
	  20.7 & & 20.5 & 20.7 & & 20.5 & 20.7 & & 20.3 \\
	  21.7 & 21.5 & 20.5 & 21.9 & 21.7 & 20.5 & 21.7 & 21.3 & 20.5 \\
	 \hline
	\end{tabular}
\end{table}

\subsubsection{Variability}

Information about the general photometric variability of sources is derived as in \cite{cross2009}. In practice, for each detection processed through the VDFS by WFAU, a variability flag is set to true (1) or false (0) using the sum of the weighted ratios of the intrinsics standard deviation to the expected noise. The weighting in each filter depends on the respective number of observations; at least five observations in one filter are needed for an object to be counted as variable. Thus, for the VMC data this is driven by observations in the $K_\mathrm{s}$ band. Mean, median, minimum and maximum magnitudes as well as rms, median absolute deviation, the probability of being variable in a given filter and other attributes are also calculated and reported in variability catalogues. Note that for periodic variables amplitudes, derived from the difference between maximum and minimum magnitudes, as well as mean values will probably differ from those obtained from fitting their light-curves with templates that cover the entire phase of variation.

\subsubsection{PSF photometry}

Each VMC tile is also accompanied by a catalogue with point-spread-function (PSF) magnitudes. The PSF detections are extracted separately in each filter following the method described in \cite{rubele2015}, then the catalogues are correlated using a radial distance threshold of 1 arcsec. This method combines the calibrated pawprint images using the SWarp\footnote{https://www.astromatic.net/software/swarp/} programme \citep{bertin2002} to generate a uniform sky subtracted deep tile image. Artefacts in the pawprint images are removed by masking contaminated regions during the co-addition. The PSF in each detector on each pawprint image is normalised to a constant PSF reference model, constructed from the largest effective PSF model among all detectors and pawprints,  before combining them. This is a sort of a homogenisation process to account for seeing variations across the tile. We refer the reader to \cite{rubele2015}, their Appendix A, for a detailed description and visualisation of the procedure. The uniformity of limiting magnitude on the final deep tile is intrinsically dependent on differences in the detector sensitivity and stellar crowding. The PSF magnitudes are aligned with the VDFS magnitudes and are not corrected for reddening. However, the name of sources following the International Astronomical Union convention\footnote{https://www.iau.org/public/themes/naming/} (IAUNAME) in the PSF catalogues may not be unique. At this stage, sources in the overlap of tiles will appear with the same IAUNAME. Furthermore, the IAUNAME is rounded to two decimal points in arcsec, hence it may be possible that two sufficiently close extractions result in two sources with the same IAUNAME. The catalogues contain parameters that link the sources, extracted with PSF photometry, with those extracted with VDFS photometry. The SOURCEID that uniquely identifies sources in the VDFS catalogues may correspond to multiple UNIQUEIDs, a UNIQUEID identifies a PSF source, but distances in arcmin to each counterpart are provided.

The SHARP parameter, which is a measure of the difference between the observed width of the object and the width of the PSF model, and STAR\_PROB parameter listed in the catalogues could be used to disentangle point-like and extended sources. For example, for stellar objects STAR\_PROB>0.77 and SHARP<0.5 whereas cosmic rays have SHARP<0. The efficiency of this selection depends on the FWHM and signal-to-noise ratio of the image. Sources that are close to saturation are not specifically flagged. The PSF photometry detects sources which are on average a few magnitude fainter than those detected with the aperture-based VDFS photometry. The magnitude difference may be larger in crowded stellar fields, especially in the $Y$ band, or smaller in less crowded fields and in the $K_\mathrm{s}$ band. 

The completeness of the catalogues is evaluated from artificial star tests and PSF photometry. The mean completeness and standard deviation among all VMC tiles, without including the observations from the additional programmes, is listed in Table \ref{psf}. This table shows for each filter the mean magnitude tracing the 80 and 50 percent fractions of artificial stars recovered, with the respective uncertainties. We refer the reader to \cite{rubele2012}, their Figs.\,4 and 5, for an illustration. The additional programmes for which one $K_\mathrm{s}$ tile observation is added to the VMC products would likely produce PSF photometric detections and completeness results that are not too different from those achieved from the VMC data alone. The PSF catalogue for tile SMC-gap does not contain the completeness information, but this tile largely overlaps with the adjacent tiles for which the completeness is available; the sources within the gap will have similar values. However, for tiles LMC 7\_5 and SMC 5\_4, for which many additional observations were obtained in the $J$ and $K_\mathrm{s}$ bands, the completeness values will be replaced in an ongoing study to characterise the young stars that includes the execution of the artificial star tests.

\begin{table}
	\small
	\setlength{\tabcolsep}{2.5pt}
	\caption{Completeness of PSF catalogues.}
	\label{psf}
	\centering
	\begin{tabular}{ccccc}
	\hline\hline
	 Filter & Mean 80\% & Uncertainty 80\% & Mean 50\% & Uncertainty 50\% \\
	\hline
	 $Y$ & 21.34& 1.21 & 22.24 & 0.97 \\
	 $J$ & 21.07 & 1.17 & 21.86 & 1.03 \\
	 $K_\mathrm{s}$ & 20.31 & 1.05 & 20.87 & 0.91 \\
	 \hline
	\end{tabular}
\end{table}

\subsubsection{Example: tile LMC 3\_3}
\begin{figure}
\centering
\includegraphics[width=\hsize]{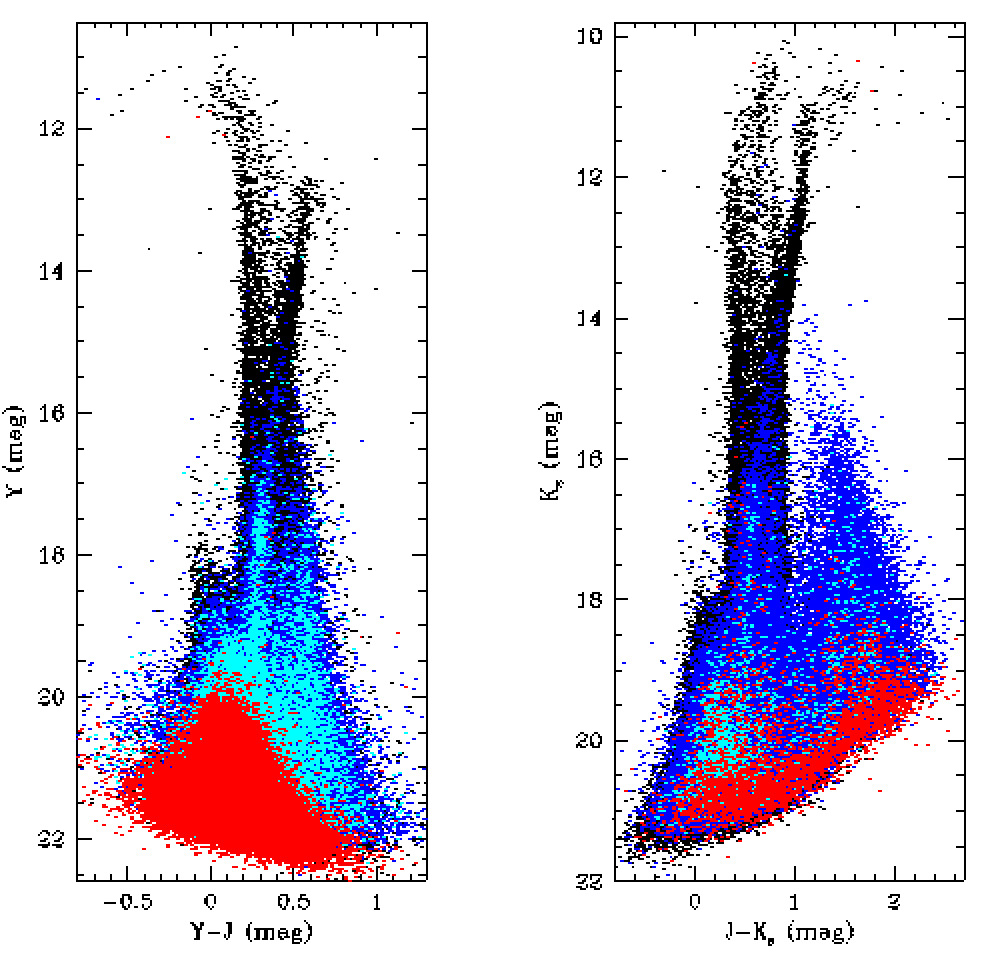}
\caption{Distribution of the VMC sources from tile LMC 3\_3 in the colour--magnitude diagrams ($Y$, $Y-J$) on the left and ($K_\mathrm{s}$, $J-K_\mathrm{s}$) on the right. Different types of detections are colour-coded as follows. Sources with a stellar or a galaxy profile which are detected in three bands are shown in black and blue, respectively. Similar sources detected only in two bands are shown in red and turquoise.}
\label{lmc33cmd}
\end{figure}

\begin{figure}
\centering
\includegraphics[width=\hsize]{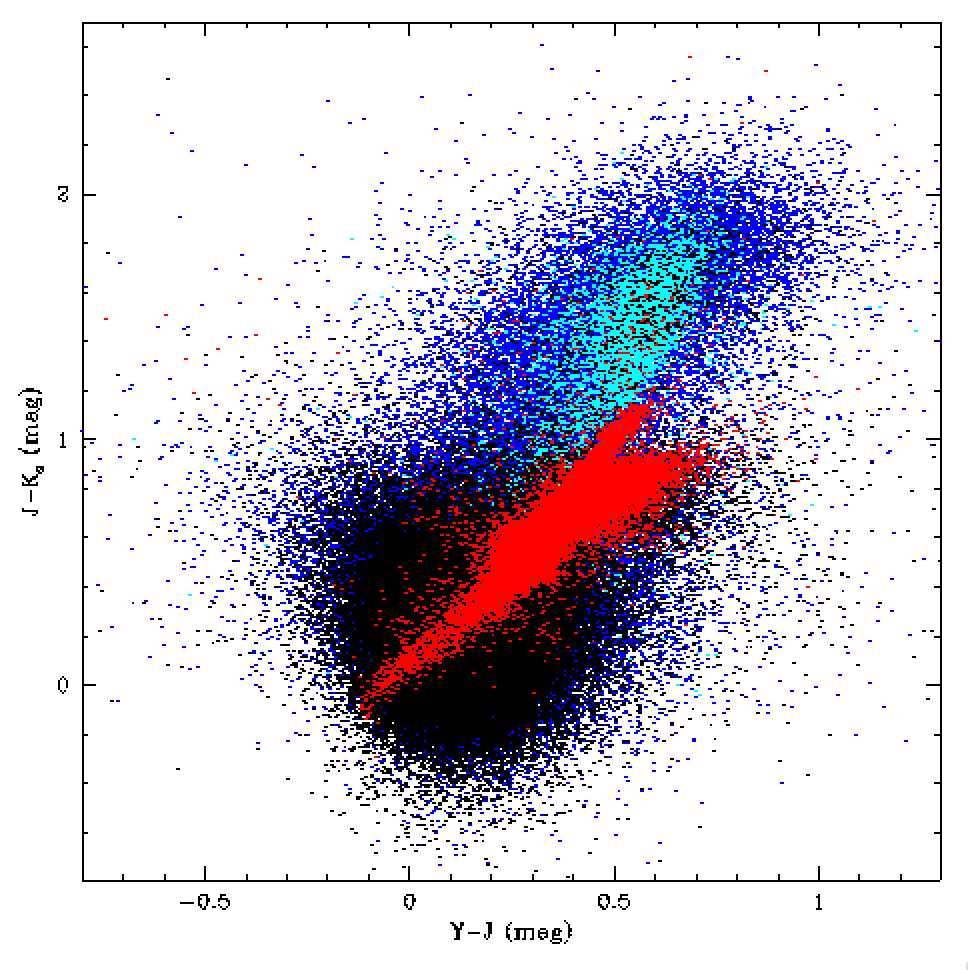}
\caption{Distribution of VMC sources from tile LMC 3\_3 in the colour--colour diagram. Sources with a stellar or a galaxy profile are shown in black and blue, respectively. Similar sources selected to have photometric uncertainties in all three bands <0.05 mag are shown in red and turquoise.}
\label{lmc33col}
\end{figure}

\begin{figure}
\centering
\includegraphics[width=\hsize]{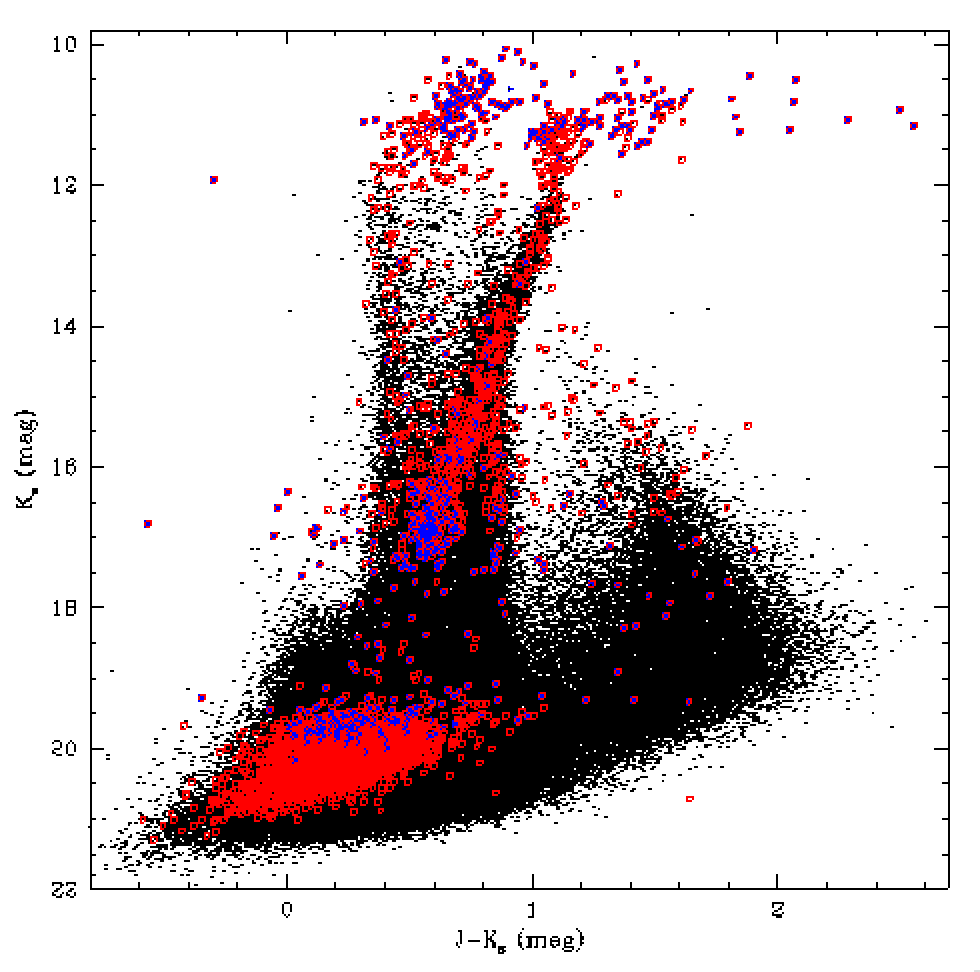}
\caption{Distribution of VMC sources from tile LMC 3\_3 in the colour--magnitude diagram $K_\mathrm{s}$, $J-K_\mathrm{s}$ with sources flagged as variable (4\,275) highlighted in red. Among them 524 (blue) are detected in three bands, have {\it rms}<1 and amplitude, defined as the difference between the maximum and minimum $K_\mathrm{s}$ magnitudes, <0.4 mag.}
\label{lmc33var}
\end{figure}

To illustrate some of the aspects mentioned in the previous subsections we focus on tile LMC 3\_3. This tile is located in the southern part of the LMC (see Fig.\,\ref{lmcmap}) in a region of moderate stellar density; tiles in the inner regions of both Clouds have about twice as many sources. Tile LMC 3\_3 contains in total about 640\,000 sources of which 55\% are detected in three bands, 25\% only in two bands and 20\% only in one band. Figure \ref{lmc33cmd} shows the distribution of these sources in the ($Y$, $Y-J$) and ($K_\mathrm{s}$, $J-K_\mathrm{s}$) colour--magnitude diagrams. Extended sources can be either stars or galaxies regardless of wether they are detected in three or two bands, whereas sources detected only in the $Y$ and $J$ bands are most likely stars. Milky Way stars have not been removed from these diagrams and we refer the reader to \cite{elyoussoufi2019} for a detailed explanation of the stellar population features through a comparison with theoretical models. Figure \ref{lmc33col} shows the distribution of sources from tile LMC 3\_3 in the colour--colour diagram. Stars and galaxies occupy clearly distinct regions (see also \citealp{cioni2013}). In this tile, about 4\,300 sources show variability in the $K_\mathrm{s}$ band, according to the criteria described in \cite{cross2009}, and among them about 500 have an amplitude larger than 0.4 mag (Fig.\,\ref{lmc33var}). Bright red giants, RR Lyrae stars, some faint stars as well as YSOs, which share their location with background sources (see \citealp{zivkov2020}; their Fig.\,11) are among these most-variable sources. 

\begin{figure}
\centering
\includegraphics[width=\hsize]{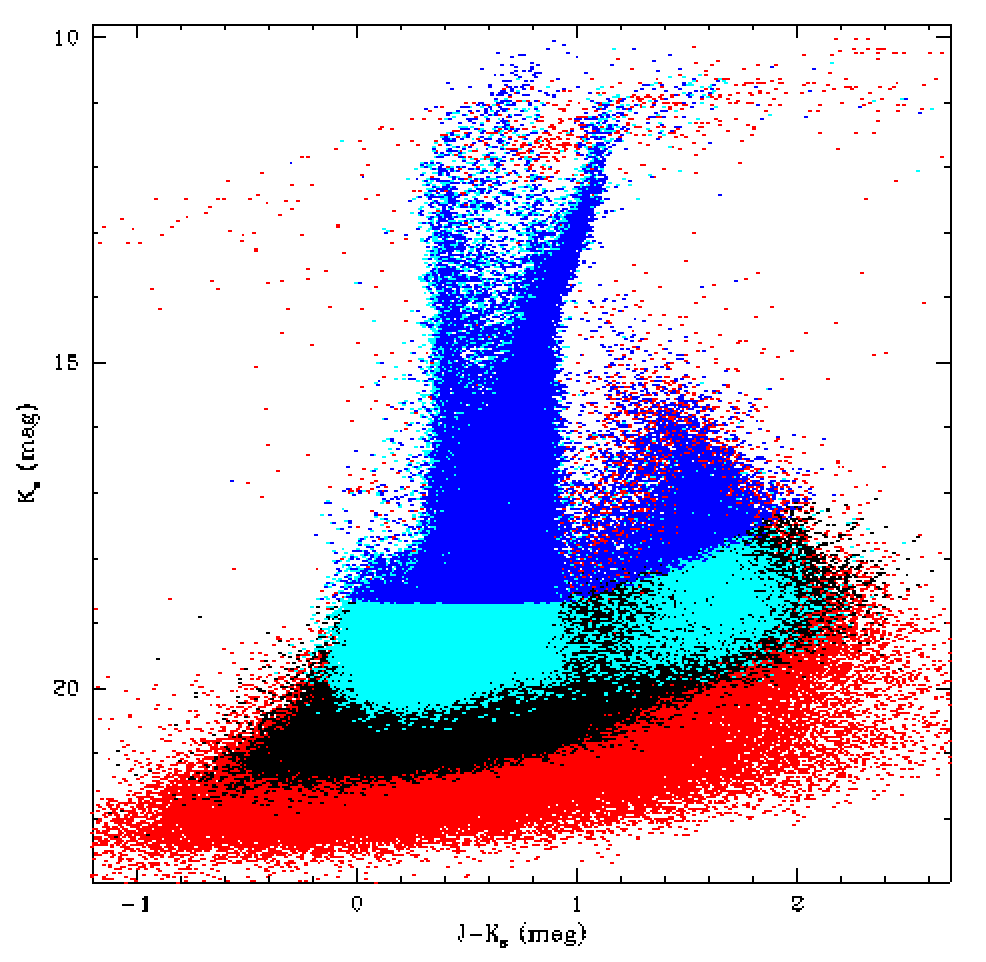}
\caption{Distribution of VMC sources from tile LMC 3\_3 in the colour--magnitude diagram $K_\mathrm{s}$, $J-K_\mathrm{s}$ colour-coded as follows. All sources with PSF photometry as shown in red while all sources with VDSF photometry are shown in black. The corresponding sources with photometric uncertainties in all three bands <0.05 mag are shown in blue and turquoise, respectively.} 
\label{lmc33psf}
\end{figure}

A comparison between sources with VDFS and PSF magnitudes is shown in Fig.\,\ref{lmc33psf}. In tile LMC 3\_3 there are about 1\,500\,000 sources with PSF magnitudes, about a factor of two more than those with VDFS magnitudes. In this tile, the PSF photometry detects sources about 2 magnitudes fainter then in the VDFS photometry. Note that for sources with PSF magnitudes and photometric uncertainties <0.05 mag there is a clear separation between stars and galaxies. Galaxies depict a triangular distribution centred at about $(J-K_\mathrm{s})=1.5$ mag whereas stars have  $(J-K_\mathrm{s})<1$ mag. At the brightest magnitudes there are PSF sources close to the saturation limit for which their magnitudes may not be reliable; they span a horizontal colour range at $K_\mathrm{s}\sim11$ mag.

\subsection{Additional data products}
\label{add_products}

\subsubsection{Proper motion tables}
In this release, stellar proper motions for about 12 million individual sources detected throughout the VMC area are made publicly available for the first time. The proper motions are those derived and analysed in \cite{vijayasree2025} which follow the same steps as in \cite{niederhofer2022}, but refer to a longer time baseline. To recap, single-epoch PSF detections are cross-matched with the deep multi-band catalogue, keeping only sources that are detected in both the $J$ and $K_\mathrm{s}$ bands. Then, only stellar sources are selected -- those must fail the background galaxies criteria of \cite{bell2019}: ($J-K_\mathrm{s}$)<1 mag, $K_\mathrm{s}$<15 mag, sharpness index for both $J$ and $K_\mathrm{s}$ filters <0.3 and stellar probability > 34\%. The proper motion of each object is calculated following a linear least-squares fit to the coordinates as a function of time, in the frame of reference of the best observed epoch. These relative values are subsequently calibrated to an absolute frame using high-quality stars in common between the VISTA sample and Gaia Data Release \#3 (DR3; \citealp{vallenari2023}) for the LMC and SMC.  Stars that (i) are not members of the LMC or the SMC according to \cite{luri2021}, (ii) occupy regions of the colour-magnitude diagram inconsistent with being Clouds stars, or that (iii) have $\sigma(K_\mathrm{s})$>0.5 mag are removed. We use instead background galaxies to calibrate stars in the Bridge and Stream because of their large number. We refer the reader to the above papers for the details on the process.

Table \ref{motion} shows the first nine lines of the proper motions for stars within tile LMC 2\_3. Columns are as follows:  $\langle\alpha\rangle$ and $\langle\delta\rangle$ provide the celestial coordinates of the stars. These are the same average coordinates present in the PSF catalogues where the average results from the combination of the coordinates in the $Y$, $J$ and $K_\mathrm{s}$ bands. $\mu_W=-\mu_\alpha \cos(\delta)$ and $\mu_N=\mu_\delta$ provide the proper motions in the West and North directions, respectively. These are followed by the scatter (rms) of the proper motion fit in both directions and the reference epoch (in decimal years) for the calculation of the proper motion. This is the epoch to which we transformed the stellar positions of all other epochs. Entire tables, one for each tile in the survey area, are only available at the Strasbourg astronomical Data Center (CDS). The proper motion values are calculated from the combination of VMC observations (Sect.~\ref{vmc}) with observations from the additional programmes (Sect~\ref{epochs}), which together correspond to 12--14 epochs and cover a time baseline of 7.00--13.25 years depending on tile (see \citealp{vijayasree2025}). They have been strictly derived for tiles that share the same centroid positions and tile patterns to minimise distortion effects introduced by the combination of different detectors. In fact, the sources located within the SMC-gap tile span a time baseline of only five years because they were treated independently from the observations of the adjacent overlapping VMC tiles. Unfortunately, the time baseline is just two years for sources located within tile SMC 5\_4 because, due to the dithering tile patterns, observations from the VMC survey and from programme 099.C-0773 do not share the same centroid position and were not combined. There are in total 110 tables corresponding to the LMC (68 tables), SMC (28 tables including the SMC-gap tile), Bridge (13 tables) and Stream (1 table) components of the VMC survey. There is no proper motion table for tile STR 2\_1 because no stars within it were identified to belong to the Clouds, using the same criteria applied to all other tiles.  

\begin{table}
	\setlength{\tabcolsep}{2.5pt}
	\caption{Stellar proper motions in tile LMC 2\_3.} 
	\label{motion}
	\centering
	\tiny
	\begin{tabular}{llrrccc}
	\hline\hline
	$\langle\alpha\rangle$ & $\langle\delta\rangle$ & $\mu_W$ & $\mu_N$ & rms$_W$ & rms$_N$ & Ref.\,Epoch \\
	(deg) & (deg) & (mas yr$^{-1}$) & (mas yr$^{-1}$) & & & (yr) \\
	\hline	
 	73.82560 & --74.96339 & 0.514 & --4.176 & 0.030 & 0.027 & 2016.78 \\
	73.82560 & --74.96339 & 5.001 & --3.448 & 0.034 & 0.034 & 2016.78 \\
	70.82896 & --74.57324 & 4.064 & 4.598 & 0.047 & 0.044 & 2016.78 \\
	73.48917 & --75.13176 & 3.311 & --2.082 & 0.070 & 0.069 & 2016.78 \\
	73.48917 & --75.13176 & 3.999 & --0.651 & 0.036 & 0.079 & 2016.78 \\
	73.68175 & --74.58508 & 3.202 & 2.804 & 0.073 & 0.084 & 2016.78 \\
	73.27473 & --74.58350 & 2.856 & --0.817 & 0.023 & 0.038 & 2016.78 \\
	73.82327 & --74.57178 & 3.639 & --0.340 & 0.042 & 0.045 & 2016.78 \\
	73.70406 & --74.56863 & 1.760 & --0.756 & 0.055 & 0.046 & 2016.78 \\
	\hline
	\end{tabular}
	\tablefoot{The proper motions are calculated with respect to the reference epoch, using 14 epoch data over a time baseline of 9.33 years. The entire table contains 39\,109 sources and is available only at CDS.}
\end{table}

\subsubsection{Parameters of variable stars}
Several tables containing the parameters of variable stars derived from the VMC data, which are already publicly available through journal publications, are included with the data release. There is a table for Cepheids (Table \ref{classical}) containing the parameters for stars located in the LMC, the SMC and part of the Bridge. The table has the following format: an identifier (ID) which for most stars comes from OGLE, the celestial coordinates of the stars ($\alpha$ and $\delta$), the pulsation mode (where F = Fundamental, 1O = First Overtone, 2O = Second Overtone and 3O = Third Overtone) and period, the single-epoch $V$-band magnitude, the intensity-averaged magnitude $\langle Y \rangle$ with its uncertainty $\sigma_Y$, the peak-to-peak amplitude $A(Y)$ with its uncertainty $\sigma_A(Y)$ and similar quantities in the $J$ and $K_\mathrm{s}$  bands. Missing values for $Y$ counterparts and amplitudes in any wave band are replaced by the value --9.9995e-8. The $E(V-I)$ colour index is also provided. The information for classical Cepheids contained in this tables is extracted and complemented from the information published in \cite{ripepi2017,ripepi2022} and we refer the reader to these publications for further details about the determination of the individual parameters. There are in total 9\,293 classical Cepheids and only data from the standard VMC programme are used for their investigation. The main difference between the SMC and LMC studies is in the usage of v1.3 and v1.5, with re-calibrated zero-points, of the VDFS. The values for classical Cepheids replace those included in DR4. 

There are also type II Cepheids and anomalous Cepheids which refer to the studies by \cite{sicignano2024,sicignano2025}. Compared to the information provided for classical Cepheids, Column 4 may list the class of type II (BLHer = BLHerculis, RVTau = RV Tauri, WVir = W Virginis, pWVir = peculiar W Virginis) or anomalous Cepheid whereas the remaining columns provide similar information (Table \ref{classical}). In total, there are 339 and 200 type II and anomalous Cepheids in the Clouds from the VMC studies.  Furthermore, the analysis of the VMC counterparts for Eclipsing Binary stars was presented in \cite{muraveva2014}. This study produced 874 sources across the LMC (Table \ref{eclipsing}), with at least 13 epochs in the $K_\mathrm{s}$ band, out of a sample of 999 detections. Table \ref{eclipsing} shows the identification from external catalogues (Catalogue), celestial coordinates, number of epochs in the $K_\mathrm{s}$ band ($N_\mathrm{Epochs}$), $K_\mathrm{s}$ magnitude at maximum light ($K_\mathrm{s}$ max) and its uncertainty ($\sigma_{K_\mathrm{s}}$ max), period, epoch of minimum light (Epoch min) and type of binary (e.g. contact or non-contact).

Quantities for RR Lyrae stars studied by \cite{muraveva2018} and \cite{cusano2021} are made available in a homogeneous way with the data release. There are in total 25 081 RR Lyrae stars with reliable VMC $K_\mathrm{s}$ mean magnitudes and amplitudes in the Clouds (22 084 in the LMC and 2 997 in the SMC). They are listed in Table \ref{rrl} (of which only a few lines are shown as an example). The reader should refer to the specific publications for values other than those included in Table \ref{rrl} and their determination. The content of Table \ref{rrl} is as follows: the VMC source ID, the celestial coordinates, the type of RR Lyrae (a, b, or c), the period, the magnitude in the $V$ band, the mean $K_\mathrm{s}$-band magnitudes with their uncertainties, the peak-to-peak amplitudes in the $K_\mathrm{s}$ band, the iron abundances ([Fe/H]) with the respective uncertainties ($\sigma_\mathrm{[Fe/H]}$) where values of --9.99995e--8 indicate that the quantities were not measured, and the $E(V-I)$ values. Four LMC sources that were listed in \cite{cusano2021} are removed here because their magnitudes appear problematic. 

Among the pulsating variables, we find the class of long-period variables (LPVs) which were investigated by \cite{gullieuszik2012} and \cite{groenewegen2020}. In this data release we include the results of the light-curve analysis as published in \cite{groenewegen2020} for 1299 sources. Table \ref{agb} shows a few example lines listing for each source: the celestial coordinates, the reduced chi-squared ($\chi^2$), the mean $K_\mathrm{s}$-band magnitudes with their uncertainties, the period and amplitude of variation in the $K_\mathrm{s}$-band with the corresponding uncertainties. Not all sources turned out to vary in the $K_\mathrm{s}$-band and for those the last four columns are left empty. We refer the reader to the original publication by \cite{groenewegen2020} for the calculation of these and other parameters.
In addition, we also include a table with the parameters of the LPV candidates (Table \ref{lpv}). The latter is a merged table obtained from both studies which provides information for  $\sim$600 sources of which 217 are red asymptotic giant branch (AGB) stars, they have colours $J-K_\mathrm{s}>3$ mag, pulsation periods $>450$ days and a spectral energy distribution (SED) typical of AGB stars. In this table we list: the celestial coordinates of the sources, the corresponding mass-loss rates log(MLR) and luminosities ($L$) as well as the provenance, 1 for \cite{gullieuszik2012} and 2 for \cite{groenewegen2020}. Some mass-loss rates were too small to be determined and the field is then left empty. In addition, seven sources from \cite{gullieuszik2012} were removed because they lack a luminosity value and two sources have a double entry, one per study. 

Candidate variable stars in a specific region of the LMC (tile LMC 7\_5) were identified by \cite{zivkov2020} through an analysis $J$ and $K_\mathrm{s}$-band light curves (independently and combined) with the scope of characterising magnitude variations in YSOs. This work produced variables across the colour-magnitude diagrams that may also be included into the groups presented above. There are about 3\,000 candidates from this study and the table produced by the authors is associated in full to this data release. Table \ref{young} shows a few example lines including: the celestial coordinates, the magnitudes in the $Y$, $J$ and $K_\mathrm{s}$ bands, and the bands in which the variability is identified. These are followed by observed amplitude of variations for the $J$ and $K_\mathrm{s}$ bands, and the type of variable according to the OGLE data base \citep{udalski2008}; in the example EB corresponds to eclipsing binary. Note that photometric uncertainties on the magnitudes are about 10 times smaller than the amplitudes in the respective bands.

\begin{table*}
	\setlength{\tabcolsep}{2pt}
	\caption{Example of the parameters of classical (9\,293), type II (339) and anomalous (200) Cepheids.} 
	\label{classical}
	\centering
	\scriptsize
	\begin{tabular}{lllcrlllllllllllllr}
	\hline \hline
	ID & $\alpha$ & $\delta$ & Mode & Period & $V$ & $\langle Y \rangle$ & $\sigma_Y$ & $A(Y)$ & $\sigma_A(Y)$ & $\langle J \rangle$ & $\sigma_J$ & $A(J)$ & $\sigma_A(J)$ & $\langle K_\mathrm{s} \rangle$ & $\sigma_{K_\mathrm{s}}$ & $A(K_\mathrm{s})$ & $\sigma_{A(K_\mathrm{s})}$ & $E(V-I)$ \\
	 & (deg) & (deg) & & (day) & (mag) & (mag) & (mag) & (mag) & (mag) & (mag) & (mag) & (mag) & (mag) & (mag) & (mag) & (mag) & (mag) & (mag) \\
	 \hline

	OGLE-LMC-CEP-0001 & 67.74050	& --69.06039 & 1O & 0.307	 & 18.154	& 17.550 & 0.003 & 0.324 & 0.014 & 17.460 & 0.002 & 0.224 & 0.010 & 17.221 & 0.014 & 0.159 & 0.053 & 0.104 \\
        OGLE-LMC-CEP-0002 & 67.94605	& --69.81931 &	F & 3.118 &	16.420 & 15.356 & 0.002 & 0.233 & 0.059 & 15.147 & 0.002 & 0.190 & 0.004 &	14.739 & 0.004 & 0.142 & 0.012 & 0.132 \\
        OGLE-LMC-CEP-0003 & 68.77376 & --70.42414 & 1O & 0.350 & 18.384 & 17.583 & 0.005 & 0.307 & 0.017 & 17.446 & 0.005 & 0.210 & 0.012 & 17.185 & 0.009 & 0.169 & 0.016 & 0.144 \\
	OGLE LMC-T2CEP-164 & 84.51801 & --70.34145 & WVir & 8.502 & 16.937 & 15.604 & 0.009 & 0.223 & 0.018 &	15.285 & 0.008 & 0.198 &	0.017 & 14.513 & 0.011 &	0.193 & 0.026 & 0.269 \\
	OGLE-LMC-T2CEP-001 & 69.56386 & --68.25889 & BLHer & 1.813 &18.440 & 17.497 & 0.032 & 0.485 & 0.155 & 17.242 & 0.008 & 0.360 & 0.056 & 16.885 & 0.009 & 0.372 & 0.021 & 0.061 \\
	OGLE-LMC-T2CEP-002 & 70.64150 & --68.61768 &	WVir & 18.326 & 16.732 &	15.282 & 0.008 & 0.859 &	0.011 & 15.114 & 0.006 & 0.797 & 0.015 & 14.506 &	0.018 & 0.758 & 0.025 & 0.117 \\
	OGLE-LMC-ACEP-112 & 79.47427 & --69.08564 & 1O & 0.693 & 18.093 & 15.068 & 0.007 & 0.033 & 0.023 & 14.466 & 0.002 & 0.030 &	0.009 & 13.404 & 0.003 &	0.015 & 0.005 & 0.121 \\
        OGLE-SMC-ACEP-072 & 16.67173 & --74.69960 & F & 1.234 & 18.068 &	14.885 & 0.007 & 0.028 &	0.012 & 14.649 & 0.005 &	0.034 & 0.014 & 14.179 &	0.005 & 0.018 & 0.009 & 0.072 \\
	OGLE-LMC-ACEP-053 & 82.77589 & --68.72923 & F & 1.888 & 17.309 & 16.375 & 0.009 & 0.413 &	0.027 & 16.043 & 0.004 &	0.390 & 0.010 & 15.684 &	0.003 & 0.266 & 0.008 &0.113 \\
       	 \hline
	\end{tabular}
	\tablebib{\cite{ripepi2017,ripepi2022} and \cite{sicignano2024,sicignano2025}.}
\end{table*}

\begin{table*}
	\caption{Example of parameters of the parameters of 999 Eclipsing Binaries.} 
	\label{eclipsing}
	\centering
	\small
	\begin{tabular}{lccclcrcl}
	\hline \hline
	 Catalogue & $\alpha$ & $\delta$ & $N_\mathrm{Epochs}$ & $K_\mathrm{s}$ max & $\sigma_{K_\mathrm{s}}$ max & Period & Epoch min & Type \\
	  & (deg) & (deg) & &(mag) & (mag) & (day) & (day) & \\ 
	 \hline
	OGLEIII &	84.29542 & --69.95295 &	14&	17.446&	0.062&	13.8688	&53599	&ch \\
	OGLEIII &	84.16158 & --69.96544 &	14&	16.167&	0.019&	30.8883	&53612	&ch	\\
	OGLEIII &	84.25529 & --69.96866 &	14&	16.855&	0.028	&18.8608	&53613&	ch \\
	OGLEIII & 83.99221 & --69.97239 &	14&	13.849&	0.004&	137.3480	&53846&	ch	 \\
	OGLEIII &	85.91154	& --69.97994 &	14&	15.492&	0.012&	61.1102&	53715&	ch	\\
	OGLEIII &	84.31954	& --69.98878 &	14	&18.620&	0.132&	0.9922	&53542&	n/c	\\
	EROS-2 &	83.66988  & --69.98929 &	14	&16.523&	0.024	&1.9908	&51917&	cont.-like	 \\
	OGLEIII &	85.28325	& --69.98890  &  14	&15.233	&0.009&	0.9776&	53511&	n/c	 \\
	EROS-2 &	83.56004	& --69.99942 &14	&16.682	&0.023&	1.0281	&51764	&non-contact\\
	\hline
	\end{tabular}
	\tablebib{\cite{muraveva2014}.}
\end{table*}

\begin{table*}
	\setlength{\tabcolsep}{2.5pt}
	\caption{Example of the parameters of RR Lyrae stars in the SMC (2\,997) and LMC (22\,084).} 
	\label{rrl}
	\centering
	\small
	\begin{tabular}{lllrlllllrrr}
	\hline \hline
	VMC\,ID & $\alpha$ & $\delta$ & Type & Period & $V$ & $\langle K_\mathrm{s} \rangle$ & $\sigma_{K_\mathrm{s}}$ & $A(K_\mathrm{s})$ & [Fe/H] & $\sigma_\mathrm{[Fe/H]}$ & $E(V-I)$ \\
	 & (deg) & (deg) & & (day) & (mag) & (mag) & (mag) & (dex) & (dex) & (mag) & (mag)  \\
	 \hline
	 558384363978   & 65.58050 & --70.65125 & c &     0.303 &   19.54 &  18.64 &  0.05 &    0.577 &  99.0 &   99.0   &    0.07    \\
          558384345675  & 65.63913 & --70.54847 & ab &    0.669 &   19.31 &  17.81 &  0.03 &    0.398 &  $-$1.84  & 0.093 &   0.10     \\
          558384349060  & 65.68250 & --70.57250 & ab &    0.565 &   19.43 &  18.19 &  0.03 &    0.420 &   $-$1.78 &  0.051 &   0.07    \\
          558384359910  & 65.85117 & --70.65383 & ab &    0.544 &   19.45 &  18.12 &   0.03 &    0.215 &  $-$1.79 &  0.058  &  0.07    \\
          558384290896  & 66.11563 & --70.29554 & ab &    0.608 &   19.51 &  18.10 &   0.03 &    0.462 &  $-$1.28 &  0.083 &   0.10     \\
          558384232839  & 66.23104 & --70.01953 & ab &   0.458 &   19.61&   18.42 &  0.04 &    0.405 &  99.0 &   99.0  &     0.03    \\
          558384211056  & 66.33921 & --69.92728 & ab &    0.597 &   19.41 &  18.00 &   0.03 &    0.431 &  $-$1.35 &  0.073 &   0.10     \\
          558384235833  & 66.45379 & --70.05522 & ab &    0.451 &   19.74 &  18.58 &  0.04 &    0.338 &  $-$1.00  &  0.043 &   0.10     \\
          558384354506  & 66.49542 & --70.68428 & c &     0.339 &   19.51 &  18.37 &  0.04 &    0.325 &  $-$1.60 &   99.0  &     0.17    \\
  	\hline
	\end{tabular}
	\tablebib{\cite{muraveva2018} and \cite{cusano2021}.}
\end{table*}

\begin{table}
	\setlength{\tabcolsep}{2.5pt}
	\caption{Example of the parameters of 1\,299 LPV light-curves.} 
	\label{agb}
	\centering
	\tiny
	\begin{tabular}{llrllrrrr}
	\hline \hline
	$\alpha$ & $\delta$ & $\chi^2 $ & $K_\mathrm{s}$ & $\sigma_{K_\mathrm{s}}$ & P & $\sigma_\mathrm{P}$ & $A(K_\mathrm{s})$ & $\sigma_{A(K_\mathrm{s})}$ \\
	(deg) & (deg) & & (mag) & (mag) & (day) & (day)& (mag) & (mag) \\
	\hline
	4.87644 &	--72.46576 &	19.8&	11.982&	0.011&	265 &	1 &	0.24&	0.05	\\
	5.99372 &	--73.63190 &	308.8&	11.473&	0.077&	479 &	6 &	0.42&	0.08	\\ 
	6.08824 &	--72.10744 &	6.7&	16.452&	0.054	 & & \\	 	 	 		 	 
	6.49831 &	--73.89577 &	268.6&	11.427&	0.081&	468 &	5 &	0.55&	0.18	\\
	6.79554 &	--73.40841 &	42.2&	11.455&	0.025&	451 &	5 &	0.32&	0.12	\\	 
	7.32971 &	--71.06385 &	863.7&	13.422&	0.227&	343 &	4 &	0.73&	0.38	\\
	7.57224 &	--72.47230 &	695.6&	10.792&	0.062&	508 &	12	&0.39&	0.18	 \\
	7.66975 &	--73.71253 &	1317.8&	11.524&	0.154&	458 &	15&	0.62	&0.52\\	 
	7.91706 &	--73.79822 &	4.8&	14.382&	0.008&	740 &	14	& 0.09&	0.05	\\
	\hline
	\end{tabular}
	\tablebib{\cite{groenewegen2020}.}
\end{table}

\begin{table}
	\setlength{\tabcolsep}{2.5pt}
	\caption{Example of the parameters of 584 LPV stars.} 
	\label{lpv}
	\centering
	\small
	\begin{tabular}{llrrc}
	\hline \hline
	$\alpha$ & $\delta$ & $\log(\mathrm{MLR})$ & $L$ & Source \\
	(deg) & (deg) & $(10^{-6}$ M$_\odot$/yr) & (L$_\odot$) & \\
	 \hline
	 4.87644 &	--72.46576 &	1.76	& 3162	 & 1 \\
	 8.11701 &	--71.78911 &	5.00	& 7943	 & 1 \\
	 8.44205 &	--72.74958 &	7.94	 & 7943	 & 1 \\
	 8.51178 &	--72.36340 &	2.51	 & 3981	 & 1 \\
	 9.16449 &	--72.27407 &	3.16	 & 5012	 & 1 \\
	 9.23627 &	--72.42153 &	3.08	& 5012	 & 1 \\ 
	 9.32881 &	--72.28424 &	1.72	& 3162	 & 1 \\
	 9.38776 &	--72.87919 &	11.22 & 5012	 & 1 \\
	 9.46602 &	--69.82997 &	2.57	& 3162	 & 1 \\
	 \hline
	 \end{tabular}
	 \tablebib{\cite{gullieuszik2012} and \cite{groenewegen2020}.}
\end{table}

\begin{table*}
	\setlength{\tabcolsep}{2.5pt}
	\caption{Example of the parameters of 3\,062 candidate variable stars.} 
	\label{young}
	\centering
	\small
	\begin{tabular}{cccccrccc}
	\hline \hline
	$\alpha$ & $\delta$ & $Y$ & $J$ & $K_\mathrm{s}$ & Variability & $A(J)$ & $A(K_\mathrm{s})$ & Variability \\
	(deg) & (deg) & (mag) & (mag) & (mag) & bands & (mag) & (mag) & type \\
	 \hline
	82.83840  & --67.36311  & 16.204&	16.000&	15.768&	J,K$_\mathrm{s}$	&0.124 &	0.163 &	  \\
	82.59136  & --67.36208 &16.757&	16.722&	16.856&	K$_\mathrm{s}$	&0.132 &	0.284 &	  \\	
	81.55287  & --67.36134	&16.903&	16.414&	15.765&	K$_\mathrm{s}$	&0.237 &	0.178 &	  \\	
	82.32587 & --67.36032 	&16.067&	15.902&	15.698&	J,K$_\mathrm{s}$	&0.100 &	0.156 &	  \\	
	82.81752  & --67.36012	&17.442&	17.020&	16.413&	J,K$_\mathrm{s}$	&0.259 &	0.357 &	  \\	
	82.59135  & --67.35977	&17.402&	17.320&	17.117&	K$_\mathrm{s}$	&0.122 &	0.412 & EB	 \\
	80.74586 & --67.35972	&17.965&	17.922&	17.321&	K$_\mathrm{s}$	&0.188 &	0.279 &	  \\	
	82.61395 &--67.35937	&17.618&	17.576&	17.705&	K$_\mathrm{s}$	&0.141 &	0.360 & EB	 \\
	82.79430 & --67.35903	&17.172&	16.601&	16.045&	J	&0.126 &	0.076 &	  \\	
	\hline
	\end{tabular}
	\tablebib{\cite{zivkov2020}.}
\end{table*}

\subsubsection{Background sources}
The VMC photometry for quasars known in the literature at the time of the study by \cite{cioni2013} and those subsequently confirmed with spectroscopic observations by \cite{ivanov2016,ivanov2024} and \cite{maitra2019} are associated with the data release. There are in total 1\,172 sources of which $\sim$300 are from VMC follow-up studies. An example of the provided information is given in Table \ref{qso}. This table contains the name of the sources, the celestial coordinates, the mean $Y$-band magnitude with its uncertainty and similar quantities for the $J$ and $K_\mathrm{s}$ bands, the VSA-Class flag (--1 = star, 1 = galaxy) and the redshift ($z$) obtained from the spectroscopic analysis. The observed spectral features and their respective wavelengths for both previously known and confirmed quasars are listed in the original publications. 

\begin{table*}
	\setlength{\tabcolsep}{2.5pt}
	\caption{Example of VMC photometry, class and redshift of 295 quasars.} 
	\label{qso}
	\centering
	\small
	\begin{tabular}{lrrccccccrc}
	\hline \hline
	ID & $\alpha$ & $\delta$ & $Y$ & $\sigma_Y$ & $J$ & $\sigma_J$ & $K_\mathrm{s}$ & $\sigma_{K_\mathrm{s}}$ & Class & $z$ \\
	 & (deg) & (deg) & (mag) & (mag) & (mag) & (mag) & (mag) & (mag) & & \\ 
	 \hline
	 VMC J001806.53-715554.2 & 4.52721 & --71.93172 & 18.236 & 0.015 & 17.933 & 0.014 & 16.392 & 0.012 & 1 & 0.620  \\ 
	 VMC J002014.74-712332.3 & 5.06142 & --71.39231 & 19.115 & 0.025 & 18.613 & 0.022 & 17.014 & 0.017 & --1 & 1.190  \\
	 VMC J002714.03-714333.6 & 6.80846 & --71.72600 & 17.766 & 0.012 & 17.439 & 0.011 & 15.905 & 0.010 &  1 & 0.474  \\ 
	 VMC J002726.28-722319.2 & 6.85950 & --72.38867 & 19.318 & 0.029 & 18.794 & 0.024 & 17.140 & 0.019 &  1 & 0.697  \\
	 VMC J002956.48-714638.1 & 7.48533 & --71.77725 & 19.216 & 0.027 & 18.847 & 0.026 & 17.832 & 0.026 & --1 & 4.098  \\
	 VMC J003430.32-715516.4 & 8.62633 & --71.92122 & 18.774 & 0.021 & 18.350 & 0.019 & 17.341 & 0.021 & --1 & 1.860  \\
	 VMC J003530.33-720134.5 & 8.87638 & --72.02625 & 19.033 & 0.025 & 18.539 & 0.022 & 17.269 & 0.020 & --1 & 0.667  \\
	 VMC J011858.84-740952.3 & 19.74517 & --74.16453 & 19.102 & 0.024 & 18.694 & 0.021 & 17.477 & 0.021 & --1 & 3.003  \\
	 VMC J011932.23-734846.6 & 19.88429 & --73.81294 & 19.257 & 0.027 & 18.807 & 0.022 & 17.170 & 0.018 & --1 & 2.143 \\
	\hline
	\end{tabular}
	\tablebib{\cite{cioni2013}, \cite{ivanov2016,ivanov2024} and \cite{maitra2019}.}
\end{table*}

\cite{bell2020,bell2022} provide the parameters of about half a million and 2.5 million extragalactic sources behind the SMC and the LMC, respectively, see also Sect.~\ref{sed}. Their tables are now included in the VMC final data release to facilitate the query and visualisation of the data products together with the other VMC data sets. Table \ref{bell} shows a few example lines giving: the celestial coordinates, the best-fitting photometric redshift ($z_\mathrm{BEST}$) with the lower ($z_\mathrm{BEST}^{-\sigma}$) and upper uncertainties ($z_\mathrm{BEST}^{+\sigma}$), similar information for the maximum-likelihood photometric redshift ($z_\mathrm{ML}$), the best-fitting galaxy (E: 1--21, Sbc: 22--37, Scd: 38--48, Irr: 49--58, Starburst: 59--62) template with the corresponding $\chi^2$, and the best-fitting $E(B-V)$. As well as the background galaxies there are also $\sim$22\,000 AGN \citep{bell2022} behind the LMC for which the same information is provided, but with the template numbers as follows: (1) Seyfert 1.8, (2) Seyfert 2, (3--5) QSO, (6--7) type-2 QSO, (8--9) Starburst/ULRIG, and (10) Starburst/Seyfert 2.

\begin{table*}
	\setlength{\tabcolsep}{2.5pt}
	\caption{Example of SED minimisation output for candidate background sources.} 
	\label{bell}
	\centering
	\small
	\begin{tabular}{llrrrrrrcrr}
	\hline \hline
	$\alpha$ & $\delta$ & $z_\mathrm{BEST}$ & $z_\mathrm{BEST}^{-\sigma}$ & $z_\mathrm{BEST}^{+\sigma}$ & $z_\mathrm{ML}$ & $z_\mathrm{ML}^{-\sigma}$ & $z_\mathrm{ML}^{+\sigma}$ & Template & $\chi^2$ & $E(B-V)$ \\
	(deg) & (deg) & & & & & & & & & (mag) \\
	\hline 
	73.29670&	--75.69781&	0.200&	0.192 &	0.208 &	0.184 &	0.150 &	0.212 &	28&	61.047&	0.00 \\
	73.77041&	--75.72749&	0.490 &	0.465 &	0.514 &	0.469 &	0.397 &	0.515 &	54&	2.105 &	0.05	\\
	73.82095&	--75.72746&	0.329 &	0.306 &	1.374 &	1.108 &	0.333 &	1.532 &	59&	8.647 &	0.00	\\
	73.78706&	--75.72451&	0.020  & 	0.020 &	0.028 &	0.013 &	0.004 &	0.031 &	49&	71.182 &	0.15	\\
	73.74649&	--75.71641&	1.028 &	0.774 &	1.218 &	0.921 &	0.529 &	1.132 &	50&	2.608 &	0.15	\\
	73.50257&	--75.71626&	0.596 &	0.535 &	1.133 &	0.924 &	0.587 &	1.258 &	57&	3.188 &	0.50	\\
	73.83151&	--75.71592&	0.486 &	0.443 &	0.539 &	0.480 &	0.398 &	0.537 &	38&	5.576 &	0.00	\\
	73.48358&	--75.71535&	1.321 &	1.311 &	1.332 &	1.330 &	1.304 &	1.364 &	62&	15.464 &	0.40	\\
	73.77131&	--75.71414&	0.056 &	0.042 &	0.067 &	0.073 &	0.042 &	2.894 &	62&	15.110 &	0.05	\\
	 \hline
	\end{tabular}
	\tablefoot{There are 2\,474\,235 (general) and 21\,828 (AGN) candidates in the LMC and 497\,577 candidates in the SMC from \cite{bell2020,bell2022}.}
\end{table*}

A probabilistic random forest supervised machine learning algorithm was used by \cite{pennock2025} to classify about 130 million VMC sources. The resulting classification is published with the data release whereas a detailed analysis of the extragalactic sources behind the Clouds is presented in the original publication. Table \ref{clara} contains the celestial coordinates of the sources and the probability of a given type of classification as: AGN (pAGN), galaxy (pGal), OB star (pOB), red giant branch star (pRGB), AGB star (pAGB), H\,{\sc ii} region or YSO (pYoung), Planetary Nebula (pPN), post-AGB or post-RGB star (pPost), red supergiant star (pRSG), high proper motion star (pHPM), likely a foreground star and Unknown source (pUnk). This is followed by the class with the highest probability (PrfClass) with the respective probability value (pPrfClass), the probability for the source to be of extragalactic nature (pExGal), the level of confidence for the classification (Flag; high: >80\% -- H, medium: between 60\% and 80\% -- M, and low: <60\% -- L) and finally the indication of whether there is an X-ray or radio counterpart (XorR) within the respective uncertainties of the complementary observations; if neither are present N is listed. The uncertainties corresponding to the different probabilities are not shown here, but are included with the data release.
	
\begin{table*}
	\setlength{\tabcolsep}{2.5pt}
	\caption{Classification of VMC sources from a probabilistic random forest supervised machine learning algorithm.} 	\label{clara}
	\centering
	\scriptsize
	\begin{tabular}{ccrrrrrrrrrrrcrrcc}
	\hline \hline
	$\alpha$ & $\delta$ & pAGN & pGal & pOB & pRGB & pAGB & pYoung & pPN & pPost & pRSG & pHPM & pUnk & PrfClass & pPrfClass & pExGal & Flag & XorR \\
	(deg) & (deg) & (\%) & (\%) & (\%) & (\%) & (\%) & (\%) & (\%) & (\%) & (\%) & (\%) & (\%) & (\%) & (\%) & (\%) & (\%) & (\%) \\
	\hline	
	9.12715  &   --70.65073 &   71.358 &    6.040 &    0.880 &     1.893 &    1.263  & 0.000 & 5.426 &  0.389	& 0.072 & 0.365 & 11.288 & AGN  &  71.358 & 77.398 & M & N   \\
        9.12716  &   --71.73568 &   10.695 &     1.139  &    0.829  &     4.997  &    6.604 & 0.000 & 7.092 &	0.449 &	0.291 &	1.128 & 66.473 &	Unk &  66.473 & 11.835 & M & N   \\
 	9.12716  &   --72.47365 &   58.366 &    5.470 &    0.091  &    3.874 &   0.905  &  0.000 &	0.699 & 0.021 &	0.018 &	0.240 &	28.977 &	AGN &  58.366 & 63.836 & M & N   \\
 	9.12716  &   --71.89797  &    42.437 &    28.003 &   0.556 &    25.191 &   1.474 &  0.000 &	2.698 &	0.201 & 0.175 & 0.353 & 20.392 &	AGN &   42.437 & 70.440 & M & N   \\
  	9.12717  &   --71.40187  &    14.386 &    0.765 &    0.072 &    15.824 &    2.076  &  0.000 & 0.032 &	0.343 & 0.191 & 1.667 & 64.567 & Unk & 64.567 & 15.151 & M & N   \\
  	9.12717 &    --70.71784  &    0.0227 &    0.008 &    0.101  &    0.068 &    0.033  & 0.000 &	0.072 &	0.011 & 0.010 & 0.013 & 0.657 & Unk & 0.657 & 0.031 & M & N    \\
  	9.12717  &   --72.56104   &   0.071 &    0.006 &   0.006 &    0.074 &    0.029  & 0.000 &	0.002 & 0.001 & 0.001 &	0.034 &	0.775 &	Unk &     	0.775 &	0.078 & M & N   \\
 	9.12717  &   --70.32687  &    0.041  &    0.017 &    0.151  &    0.065 &    0.023 & 0.000 &	0.055 &	0.017 &	0.001 &	0.009 &	0.620 &	Unk &      	0.620 & 	0.058 & M & N    \\
  	9.12718  &   --72.88727   &   0.127 &   0.011 &    0.002 &    0.069 &    0.002  &  0.000 &	0.002 &	0.001 & 0.001 &	0.002 &	0.782 &	Unk &      	0.782 &	0.138 & M & N   \\
			\hline
	\end{tabular}
	\tablefoot{Uncertainties associated to the above probabilities \citep{pennock2025} are included with the data release.}
	\end{table*} 
		
\subsubsection{Reddening}
Interstellar reddening measured towards individual red clump stars by \cite{tatton2021} is included in the data release. This is provided in the form of colour excesses $E(Y-K_\mathrm{s})$ for 561\,813 stars in the SMC (excluding those in the VISTA tile containing the 47 Tucanae globular cluster). Similarly, interstellar reddening values are provided for 2\,356\,052 red clump stars in the LMC for the first time, which are described and will be used in a forthcoming paper on the LMC's structure. These values are given with respect to the average intrinsic colour of a red clump star derived from isochrones (\citealp{marigo2008}), which corresponds to 0.76 mag in the SMC and 0.84 mag in the LMC. This produces a tail of negative extinction values, which agrees with them being random observational errors of values measured in low-extinction regions. The reddening values can be converted to $A_V$ via $E(Y-K_\mathrm{s})$=0.2711$\times A_V$ (assuming $R_V$=3.1 and the \citealp{cardelli1989} extinction law) as explained in \cite{tatton2021}. Table \ref{reddening} shows a few example lines containing the celestial coordinates, where negative values for $\alpha$ correspond to 360--$\alpha$ deg, the $E(Y-K_\mathrm{s})$ value and the same value, but obtained from the average of the values from the nearest 1\,000 red clump stars (effectively, a smoothed reddening), $E(Y-K_\mathrm{s})$ smoothed.

Towards the LMC tile containing the 30 Doradus region (VMC tile LMC 6\_6; Fig.\,\ref{lmcmap}), reddening values are provided also in the form $E(J-K_\mathrm{s})$ for 150\,328 sources. They can be converted to $A_V$ via $E(J-K_\mathrm{s})$=0.16237$\times A_V$, as in \cite{tatton2013}. Due to the high stellar density the number of sources and the corresponding mean, median and maximum extinction within 30 arcsec, 1 arcmin and 5 arcmin are also provided. Table \ref {reddening30} shows a few lines as an example (see \citealp{tatton2013} for details). 

Reddening values in the form of $E(B-V)$ are provided towards candidate background galaxies from the studies by \cite{bell2020,bell2022}, see Table \ref{bell}.

\begin{table}
	\setlength{\tabcolsep}{2.5pt}
	\caption{Example of stellar reddening for red clump stars.}
	\label{reddening}
	\centering
	\small
	\begin{tabular}{rrrc}
	\hline\hline
	$\alpha$ & $\delta$ & $E(Y-K_\mathrm{s})$ & $E(Y-K_\mathrm{s})$ smoothed \\
	(deg) & (deg) & (mag) & (mag) \\
	\hline	
	6.838300 & --75.840805 & --0.073 & 0.072 \\
	2.466070 & --75.696892 & 0.335 & 0.060 \\
	7.518070 & --75.852303 & --0.007 & 0.070 \\
	3.732880 & --75.743797 & 0.020 & 0.052 \\
	7.059740 & --75.841400 & 0.015 & 0.070 \\
	5.383160 & --75.796997 & 0.021 & 0.049 \\
	7.811191 & --75.855301 &	 0.034 & 0.069 \\
	4.930480 & --75.780098 & 0.075 & 0.046 \\
	3.117900 & --75.711700 & --0.018 & 0.057 \\
	\hline
	\end{tabular}
	\tablefoot{There are 561\,813 sources in the SMC \citep{tatton2021} and 2\,356\,052 in the LMC.}
\end{table}

\begin{table*}
	\setlength{\tabcolsep}{2.5pt}
	\caption{Example of stellar reddening for red clump stars in tile LMC 6\_6.}
	\label{reddening30}
	\centering
	\tiny
	\begin{tabular}{llrrccccrccccrcrrr}
	\hline\hline
	$\alpha$ & $\delta$ & $E(J-K_\mathrm{s})$ & N.5 & $\sigma_\mathrm{N.5}$ & M.5 & Mdn.5 & Max.5 & N1 & $\sigma_\mathrm{N1}$ & M1 & Mdn1 & Max1 & N5 & $\sigma_\mathrm{N5}$ & M5 & Mdn5 & Max5 \\
	(deg) & (deg) & (mag) & & (mag) & (mag) & (mag) & (mag) & & (mag) & (mag) & (mag) & (mag) & & (mag) & (mag) & (mag) & (mag) \\
	\hline	
	84.4710&	--70.1051	&0.158&	1&	0.000&	0.158&	0.158&	0.158&	6&	0.047&	0.111&	0.095&	0.180&	127&	 0.107&	0.162&	0.157&	0.488\\
	83.1515&	--70.1024	&0.011	&1&	0.000&	0.011&	0.011&	0.011&	9&	0.117&	0.095&	0.069&	0.385&	170&	 0.059&	0.079&	0.081&	0.385\\
	84.7594&	--70.1042	&--0.040& 2&	0.195&	0.098&	0.236&	0.236&	6&	0.117&	0.067&	0.100&	0.236&	118& 0.086&	0.106&	0.120&	0.274\\
	84.7324&	--70.1043	&0.139&	5&	0.070&	0.090&	0.137&	0.144&	9&	0.070&	0.106&	0.137&	0.193&	117& 0.080&	0.110&	0.112&	0.314\\
	84.6151&	--70.1046	&0.038&	1&	0.000&	0.038&	0.038&	0.038&	6&	0.066&	0.104&	0.099&	0.219&	110&	0.102&	0.129&	0.104&	0.391\\
	83.0735&	--70.1014	&0.079&	1&	0.000&	0.079&	0.079&	0.079&	6&	0.043&	0.045&	0.049&	0.087&	185&	0.058&	0.095&	0.092&	0.293\\
	83.8441&	--70.1048	&0.323&	3&	0.177&	0.118&	0.022&	0.323&	8&	0.215&	0.128&	0.066&	0.567&	153&	0.077&	0.059&	0.052&	0.567\\
	84.6365&	--70.1044	&0.140&	2&	0.013&	0.149&	0.159&	0.159&	8&	0.082&	0.105&	0.140&	0.208&	122&	0.095&	0.113&	0.099&	0.391\\
	85.7795&	--70.0978	&0.069&	2&	0.042&	0.039&	0.069&	0.069&	4&	0.039&	0.013&	0.010&	0.069&	81&	0.042&	--0.002&	--0.007&	0.130\\	
	\hline
	\end{tabular}
	\tablefoot{The entire table contain 150\,328 sources \citep{tatton2013}.}
\end{table*}

\subsection{Data availability}
\label{public}

All data products are stored in the VISTA Science Archive\footnote{http://vsa.roe.ac.uk} (VSA; \citealp{cross2012}). This also includes data products produced outside the VDFS, such as PSF photometry and quantities derived from the analysis of the VMC data. They are publicly accessible by the community through the VSA and the ESO Science Archive Facility\footnote{http://archive.eso.org/cms.html} (SAF; \citealp{romaniello2023}). At the VSA, the detections are organised in four main tables:\,{\it vmcDetection} for individual pawprint and tile measurements, {\it vmcSource} for band-merged catalogues from the deepest stacks, {\it vmcSynoptic} for multi-epoch observations, and {\it vmcVariability} for photometric variability statistics. There are also tables created by the VMC team for VMC specific products such as the PSF photometry ({\it vmcPsfSource}), proper motions ({\it vmcProperMotionCatalogue}), and various types of variable stars. In addition, there are cross-neighbouring tables between VMC and other survey catalogues, e.g. 2MASS \citep{skrutskie2006}, SAGE \citep{meixner2006}, OGLE \citep{udalski2015}, and Gaia \citep{prusti2016}. Due to some of the new specific features of the VMC survey, we have created a guide to using the VMC\footnote{http://vsa.roe.ac.uk/vmcGuide.html} products where we give more details and importantly Structured Query Language examples of how to use the data, with emphasis on the new team-generated tables. 

At ESO, the VMC DR7 is assigned to the VMC ESO Phase3 collection\footnote{https://doi.eso.org/10.18727/archive64/}. It is accompanied by a data-release-description file which contains further technical information, such us the nomenclature and format of the tables, which are not included in this paper. Historically, the DR1 (2011) covered only two tiles LMC 6\_6 (containing 30 Dor) and LMC 8\_8 (containing the South Ecliptic Pole) with complete VMC-survey data processed with an early version of the VDFS. Only VDFS products from both CASU and WFAU were released at that time. There was no public DR2. DR3 (2015) and DR4 (2017) added 7 and 12 tiles, respectively. The VMC data for these tiles were processed with version 1.3 of the VDFS; catalogues with PSF photometry and the parameters of classical Cepheids, Eclipsing Binaries and RR Lyrae stars across the SMC were also released. Subsequently, all VISTA data were reprocessed with version 1.5 of VDFS. This data was used in DR5 (2019) which included only VMC data products from CASU across the SMC, Bridge and Stream tiles and in DR5.1 (2019) which, for the same tiles, released the corresponding WFAU products and the PSF photometry. In 2022 DR6 provided both CASU and WFAU data products for all LMC tiles, catalogues with PSF photometry and parameters of RR Lyrae stars. The DR7 corresponds to the same CASU products released in DR5 and DR6 for the VMC programme. The CASU products for the additional programmes (Sect.\,\ref{addition}) are newly added. The WFAU data products have been reprocessed, combining the VMC-survey data with data from the additional programmes and including revised image-quality criteria. Tables with the parameters of different types of sources (e.g., stellar proper motions and redshifts of background galaxies) are also newly added. Only the tables with the PSF photometry and for the variable stars previously published remain unchanged (except for a few problematic sources which have been removed; see Sect.\,\ref{add_products} for details). We refer the reader to the CASU web pages for details about the different versions of the software and to the ESO web pages for the content of the data releases.

\section{Summary of VMC results}
\label{summary}

The main goals of the VMC survey were the determination of the spatially resolved SFH and the construction of a 3D map of the Clouds. To reach these goals, the VMC survey was designed to detect sources near the main-sequence turn-off of the oldest stellar population of the Clouds and to measure accurate mean magnitudes of pulsating variable stars, Cepheids and RR Lyrae stars, through multiple observations in the $K_\mathrm{s}$ band (aiming at <0.1 mag uncertainties). For reference, a 10-Gyr old population in the SMC has a turn off at $K_\mathrm{s}$$\sim$21 mag, which is about 0.5 mag brighter in the LMC. However, the high quality of the data enabled several additional studies. All these results are summarised below and they include both numeral consortium papers, those that are mostly based on VMC data, and papers where the VMC data complement or support other projects. Most of these studies have used previous VMC data releases, including VMC-survey data only or data from specific additional programmes, and have provided on one side an overview of the type of science that is possible with the data, and on the other they have allowed to validate the quality of the data.

\subsection{Star formation history}

The first results on the SFH in three VMC tiles covering low density regions in the LMC were presented in \cite{rubele2012}. This work demonstrates the higher depth and spatial resolution of the VMC data, compared with previous surveys in the near-infrared domain by deriving the SFH in sub-regions of 0.12 deg$^2$, together with distance and extinction. It is based on the reconstruction of colour--magnitude diagrams using stellar evolution models and it shows that by fitting a disc geometry to the galaxy, the systematic uncertainties on the star formation rate and age--metallicity relation are significantly reduced. Most sub-regions show two peaks in the star formation rate at 2 and 5 Gyr with more variations at young ages than at old ages, whereas the age-metallicity relation does not appear to vary across sub-regions. \cite{mazzi2021} presented a homogeneous analysis of 63 out of 68 VMC tiles covering the LMC. They show SFH maps with a similar spatial resolution and a resolution in log(t/yr) of 0.2--0.3 dex. They adopt a reference age--metallicity relation and adjust it by shifting to reach a best-fit solution. The galaxy appears to have formed stars at a rate of 0.3 M$_\odot$ yr$^{-1}$ between 0.5 and 4 Gyr ago, reducing to half of that value outside this range, with peaks at about 0.8 and 2 Gyr predominantly concentrated in the bar and spiral arms. The star formation at $\sim$10 Gyr encompasses instead a thick, somewhat round inner structure that does not yet resemble a bar. 

The SFH from 10 VMC tiles distributed across the main body and Wing of the SMC is described in \cite{rubele2015}. In this work, maps of the star formation rate and the total stellar mass formed at a given age have a spatial resolution of 20 arcmin. They show that the Wing formed <0.2 Gyr ago and that the SMC bar experienced a peak of star formation at $\sim$40 Myr ago. Enhanced star formation 1.5 Gyr ago is followed from the possible accretion of metal-poor gas, as revealed by a decline in the age-metallicity relation, whereas the strongest mass assembly process occurred 5 Gyr ago. A more complete picture of the SFH across the SMC was presented in \cite{rubele2018} where 14 out of 27 adjacent tiles are analysed, using improved photometric zero-points and stellar models than in the previous studies. In this work the spatial resolution is 0.143 deg$^2$ and the galaxy formed most of its mass (80\%) during the period more than 3.5 Gyr ago. A transition between a round and elongated mass distribution occurred between 5 and 3.5 Gyr ago. The Wing, the northern and southern bar regions appear as three separate structures since about 60 Myr ago. A slow chemical enrichment is confirmed from about 1 to 0.1 Gyr ago when it rises again commencing at the north-western edge of the elongated bar-like structure, to within its southern and then northern overdensities.

\subsection{Morphology maps}

The stellar evolution models inform the distribution of the colour--magnitude diagram from stars with different parameters. \cite{elyoussoufi2019} presented a morphological characterisation of the distribution of stellar populations with different median ages at a spatial resolution of 0.13 and 0.16 kpc for the LMC and SMC, respectively. These maps reveal clear substructures at intermediate ages whilst tracing the typical irregular distribution of young stars in the bar, spiral arms and tidal features and a more regular and extended distribution for the old stars. More recently, \cite{pennock2025} showed stellar population maps obtained from the application of a machine learning-based classification algorithm to the VMC data. In this work, the training set is made of a collection of AGN spectra together with spectra of galaxies and of a range of stellar classes. It yields an accuracy of at least 80\% for about a million sources (of which 2/3 in the LMC and 1/3 in the SMC) represented in the training set.

\subsection{Structure of the Clouds from variable stars}

The first results for classical Cepheids were presented in \cite{ripepi2012}. This work is mostly focused on the LMC tile containing the 30 Doradus star-forming region. It shows that the precise mean $K_\mathrm{s}$ magnitudes combined with optical light curves from large-scale monitoring programmes, like the Optical Gravitational Lensing Experiment (OGLE; \citealp{soszynski2024}), allow us to derive period--Wesenheit and period--luminosity--colour relations with a small dispersion ($\sim$0.07 mag). These empirical relations, which use for the first time the ($V-K_\mathrm{s}$) colour and time series $K_\mathrm{s}$ photometry, represent an excellent tool to measure distances and derive the structure of the galaxies. In \cite{ripepi2014,ripepi2015} a similar analysis, based on the spline interpolation of the light curves, was extended to anomalous and type II Cepheids within about a dozen LMC tiles. These are metal-poor pulsating stars contrary to classical metal-rich Cepheids. The potential of using not only Cepheids but also RR Lyrae stars (which are also metal poor) and binaries to study the 3D geometry of the Clouds was shown in \cite{moretti2014} and \cite{muraveva2014,muraveva2015} whereas new Cepheids, located in the outer region of the SMC, were discovered using only the VMC data by \cite{moretti2016}. 

Modelling simultaneously the light curve of a pulsating variable star, from visual to near-infrared (and, when available, radial velocity curves), with a non-linear convective hydrodynamical code enables us to derive distance, mass and luminosity. A sample of about 30 classical Cepheids in the Clouds analysed with these models shows that the mean distances for both the LMC and the SMC agree with literature determinations (they carry a dispersion of $\sim$0.1 mag). Moreover, the inferred masses and luminosities seem to suggest a mildly non canonical mass-luminosity relation, thus invoking the efficiency of core overshooting, and/or mass loss and/or rotation \citep{marconi2017,ragosta2019}.

The structures of the SMC, including the part of the Bridge closest to the Wing, and of the LMC were derived in \cite{ripepi2017,ripepi2022} using $>$4\,000 classical Cepheids. In these studies, intensity-averaged mean magnitudes and pulsation amplitudes are derived through the design and application of light-curve templates for modelling the VMC multi-epoch data. 
In the SMC, the Cepheids show an overall elongated distribution with younger ($\sim$120 Myr) and older ($\sim$220 Myr) stars depicting different geometries. In particular, there is an overabundance of younger stars in the north-east of the galaxy possibly resulting from a star forming episode influenced by the dynamical interaction with the LMC ($\sim$200 Myr ago), which pulled matter out of the galaxy. 
In the LMC, the spatial distribution of classical Cepheids shows features that can be explained by the dynamical interaction with both the SMC and the Milky Way. These are: a non-planar distribution with two parts of the bar displaced by $\sim$1 kpc from each other and a flared/thick disc. Furthermore, the calculated viewing angles of the bar and disc differ and the stars can be traced to two main episodes of star formation at $\sim$90 and $\sim$160 Myr ago. The relative distance modulus between the SMC and the LMC, as measured from the classical Cepheids, is $\sim$0.55 mag \citep{ripepi2016}. The empirical period--luminosity relations derived in this work include for the first time the $Y$ band and are also calculated for fundamental, first and second overtone pulsation modes. 

The structure of the SMC derived from about 3 000 RR Lyrae stars, resulting from distances measured combining OGLE IV visual light curves with intensity-averaged VMC $K_\mathrm{s}$-band magnitudes, was presented in \cite{muraveva2018}. These stars trace an ellipsoidal distribution with an average depth of 4.3 kpc. In the LMC, there are ten times more RR Lyrae stars than in the SMC and their 3D distribution, derived from a similar analysis, is also ellipsoidal. It has a similar average depth and no particular associated substructure or metallicity gradient \citep{cusano2021}. In this case, the metallicity ([Fe/H]) is obtained from the Fourier parameters of the light curves and a calibration tied to spectroscopic observations (see \citealp{skowron2016} for details).
A comprehensive study of type II Cepheids across the LMC and SMC produces a sample of $\sim$320 stars \citep{sicignano2024} for which distances agree with those from other Population II indicators. An analysis of the overall population of anomalous Cepheids across the galaxies (200 sources) shows that they also are a reliable distance indicator \citep{sicignano2025}.

\subsection{Structure of the Clouds from red giant stars}

The luminosity of red clump stars is a popular standard candle and several multi-wavelength studies in the literature characterise the structure of the Clouds with it (see \citealp{rathore2025} and references therein). Using the VMC data, \cite{subramanian2017} found that a tidal feature $\sim$11 kpc in front of the SMC is already evident 2--2.5 kpc from the centre of the galaxy.  A comprehensive study of the SMC structure using red clump stars is presented in \cite{tatton2021} who shows that the side of the galaxy nearest to the LMC exhibits the largest spatial distortions, corroborating the role played by the dynamical interaction between the two galaxies. A dust-reddening map of the SMC is also provided. A similar study of the LMC is ongoing and a reddening map of the VMC tile including 30 Doradus has already been made available in \cite{tatton2013}. This region contains $\sim$$150\,000$ red clump stars which probe reddening up to $A_V$=6 mag.

The brightness of the tip of the RGB is also a frequently used distance indicator and \cite{groenewegen2019} provided a map of the distance modulus to the Clouds based on this feature. The overall gradient across the western part of the LMC, the Bridge and the SMC is consistent with that provided by other distance indicators. Towards specific lines of sights, the method is robust with systematic errors on the distance modulus of approximately 0.045 mag and random errors better than 0.03 mag, but requires at least 100 stars in the 0.5 magnitude bin below the tip.

In \cite{choudhury2020,choudhury2021}, the slope of the RGB (in the VMC $Y$ versus $Y-J$ colour--magnitude diagram) was used as an indicator of the average metallicity across the Clouds. The spatial resolution of these analyses varies to ensure that each fitting region contains at least 60 stars. The metallicity distributions obtained from selecting good quality slopes and correlation coefficients, calibrated with respect to spectroscopic measurements, produce the following results. Both galaxies present unimodal distributions with means at [Fe/H]$=-0.97\pm0.05$ dex and $-0.42\pm0.04$ dex for the SMC and LMC, respectively. Their radial metallicity gradients are shallow and asymmetric: $-0.0031\pm0.005$ dex deg$^{-1}$ (out to $\sim$2.5 deg from the SMC centre) and $-0.008\pm0.001$ dex kpc$^{-1}$ (out to 6 kpc from the LMC centre). Towards the Bridge the gradients appear flatter than elsewhere in the galaxies. The LMC bar could also play a role in flattening the gradient in the central 3 kpc. However, since the stellar population is older than 1 Gyr, radial migration and dynamical interactions have probably also influenced the shape of the gradients.

Using a sample of $\sim$30\,000 VMC stars, mostly AGB stars and RSGs, with 3D kinematic information from Gaia DR3, \cite{kacharov2024} constructed equilibrium dynamical models to interpret the structure of the LMC. The resulting disc flattening, inclination and orientation agree with values from previous studies, and also confirms the velocity deviations from axisymmetry, especially for the young stars. The bar, which is treated as a triaxial component, has a size of $\sim$2.2 kpc, a co-rotation radius of 10 kpc and a pattern speed of 11 km s$^{-1}$ kpc$^{-1}$. This study also predicts that the central 6.2 kpc of the galaxy contains about 1.4 $\times$10$^{10}$ M$_\odot$ whereas the virial mass of the LMC as a whole corresponds to 1.8 $\times$ 10$^{11}$ M$_\odot$.

\subsection{Structure from young (non variable) stars}

Upper main-sequence stars observed by the VMC survey correspond to stellar populations younger than 1 Gyr and are useful tracers of hierarchical structures, probably related to a process of hierarchical star formation. \cite{sun2017a} identified groups of structures from several parsecs to more than 100 pc in size after computing the surface density of upper main-sequence stars. They construct a dendrogram to illustrate the nesting of the structures and compute the index of the power-law fit to their cumulative size distribution. This fractal dimension has the value of $-1.6\pm0.3$ in the 30 Doradus star forming complex \citep{sun2017a}, which is consistent with the values obtained in the LMC bar \citep{sun2017b}, the entire SMC \citep{sun2018} and LMC \citep{miller2022}, the latter hosting structures as large as 1 kpc. The mass distribution of the individual structures follows also a power law, whereas their surface density follows a log-normal distribution. The similarity of these results with those obtained from the analysis of structures in the interstellar medium supports a scenario of hierarchical star formation regulated by supersonic turbulence. By further analysing the structures with respect to their average age, it appears that the young substructures disperse within 100 Myr \citep{sun2017b}. There are overall about 600 young substructures in the SMC \citep{sun2018} and nearly 3\,000 in the LMC \citep{miller2022}.
 
The VMC sensitivity limit allows us to identify pre-main sequence populations (structures) up to an age of $\sim$10 Myr for cluster masses exceeding 1\,000 solar masses. Within one VMC tile, located just above the LMC bar, over 2\,000 such candidates were identified and characterised by \cite{zivkov2018}. Their spatial distribution clusters along ridges and filaments, with the lowest mass sources located preferentially at the outskirts of the star forming complexes. About 20\% of the VMC counterparts to known YSOs, including those associated with the pre-main-sequence structures, display aperiodic variations and are classified as eruptive, fader and dipper, with a few short-term and long-period (periodic) variables based on their VMC $J$ and/or $K_\mathrm{s}$-band light-curve. Their properties are consistent with those from Galactic studies \citep{zivkov2020}. A new method, based on a probabilistic random forest algorithm to identify and classify pre-main sequence stars with sub-solar masses (<0.5 M$_\odot$) is in preparation. This method combines near-infrared data from VISTA and optical data from SMASH. The main goal of this project is to characterise the temporal and spatial progression of star formation within two regions of $\sim$1.5 deg$^2$ (a VISTA tile) in the LMC and SMC, respectively, that encompass the most active star formation in the Clouds \citep{dresbach2024}.

\subsection{Proper motions}

The study of stellar motions through the calculation and analysis of proper motions from multi-epoch VMC data is described in several papers. \cite{cioni2014} showed the potential of the VMC data to measure proper motions with and without using data from 2MASS spanning a time baseline of one or ten years, respectively. The first proper motions based solely on the VMC data for 47 Tuc and SMC field stars were presented in \cite{cioni2016}. Note that these studies were based on stellar positions obtained from the VISTA pipeline catalogues. Subsequently, \cite{niederhofer2018a} recomputed the proper motion of 47 Tuc improving the accuracy by deriving the centroid positions using PSF photometry and observations over an extended time baseline. This technique was then applied to the entire SMC and LMC footprints as observed by the VMC survey. 

The resulting internal kinematic of the SMC, based on a sample of about two million stars, shows: (i) an outward motion suggesting a stretching of the galaxy or stripping of its outer regions, (ii) a Northward motion, possibly related to the Counter Bridge, and (iii) a coordinated motion away from the galaxy towards the Bridge and the Old Bridge \citep{niederhofer2018b,niederhofer2021}. In the central regions of the LMC, a sample of over six million stars depicts a motion where the intermediate-age/old population follows elongated orbits parallel to the bar's major axis, while the young population moves along a filamentary structure (\citealp{niederhofer2022}). In the outer regions, where the number of Milky Way stars is larger than that of the Clouds, a flow motion from the SMC to the LMC was measured (after decontamination) across the full length of the Bridge (\citealp{schmidt2020}). In addition, \cite{schmidt2022} used a machine-learning algorithm to construct a clean sample of Clouds stars and found that the slow rotational speed in the south eastern region of the galaxy may be due to the presence of SMC stars with a counter-rotating motion. By including extra observations (from the additional programmes), which increase the accuracy of the proper motions through an extended time baseline, \cite{vijayasree2025} quantify a motion away from the LMC in the northeast and southwest regions. They attribute this motion to dynamical interactions and corroborate the presence of a known and stable spatial substructure in the disc plane. In their maps, the drag of stars older than 1 Gyr towards the inner region of the bar might be a result of a relatively recent bar-formation. The analysis of the proper motions within the SMC that makes use of additional recent observations is ongoing.

\subsection{Background galaxies}
\label{sed}

The combination of the VMC photometry and the photometry at other wavelengths allows us to study the SED of stellar and extragalactic sources. In particular, a fit of the SED of AGB stars (from the optical to the near- and mid-infrared) with dust radiative transfer models provides a chemical classification of the stars into C- or O-rich and a measure of their mass-loss rate (\citealp{gullieuszik2012}). A dedicated study of AGB stars with pulsation periods >450 days, using $K_\mathrm{s}$-band magnitudes from VMC and literature data, provided mean magnitudes, periods and amplitudes for about a thousand stars \citep{groenewegen2020}. The subset of Mira-type variables is then characterised through the SED analysis; separate period--luminosity relations are constructed for C- and O-rich AGB stars. Further stellar parameters, such as the initial mass, are provided  for both the SMC and the LMC AGB population from a comprehensive calibration of the stellar models using the $K_\mathrm{s}$-band luminosity functions (\citealp{pastorelli2019,pastorelli2020}). 

In \cite{bell2019} the SEDs of background galaxies are used to estimate the redshift and classify the galaxy type based on a set of templates used with the LePHARE\footnote{http://cesam.lam.fr/lephare/lephare} $\chi ^2$-minimisation code. The SEDs are constructed combining VMC data with optical data from SMASH (\citealp{nidever2017}). The resulting reddening towards early-type galaxies (galaxies that are less intrinsically dusty than the late, more evolved, types), provides a measure of the dust intrinsic to the Clouds, assuming that the influence of the Milky Way and that of the dust outside the Clouds are properly accounted for. This pilot study shows a reasonable agreement with the dust maps available in the literature using different tracers. A similar technique is then applied to the entire SMC and LMC footprints in common between the two surveys in \cite{bell2020,bell2022}. The resulting SMC reddening map appears in good agreement with maps obtained using young stars as tracers. In the LMC a significant fraction of the dust appears located behind the bar and the best agreement is instead with maps obtained using red clump stars or derived from the SFH. These studies provide also the first categorisation of about 0.5 and 2.5 million galaxies behind the SMC and LMC, respectively.

 The combined VMC and SMASH data are also used to find galaxy clusters and groups \citep{craig2021} behind the Clouds. Maps of candidates at different redshift slices are in preparation (Craig et al., in preparation).

\subsection{Emission-line objects}

The ($J-K_\mathrm{s}$) versus ($Y-J$) colour-colour diagram (Fig.\,\ref{lmc33col}) is dominated by both the Clouds and the Milky Way stars, forming an ant-like structure (see \citealp{cioni2011} for details). The high sensitivity of VMC allows to detect extended and semi-extended objects like background galaxies, emission-line objects like PNe and quasars as well as dust-embedded YSOs. They populate low-density regions of the diagram with a minimal overlap with the dense stellar loci. VMC data are used to characterise PNe and identify possible mimics (e.g, H\,{\sc ii} regions, emission-line stars) in \cite{miszalski2011b}. This study resolves for the first time the nebular morphology of five PNe and finds candidate symbiotic stars. The discovery of a PN with a Wolf-Rayet central star demonstrates the potential of the VMC survey to find underrepresented sources \citep{miszalski2011a}.

The VMC data are also used to identify quasars behind the Clouds (\citealp{cioni2013}) using a combination of the VMC colours and the slope of the magnitude variation in the $K_\mathrm{s}$ band. Candidates, that have subsequently been spectroscopically confirmed (\citealp{ivanov2016,ivanov2024}), span a redshift $z$ between 0.1 and 4. This identification method has a success rate above 75\%. Among the candidates we also find AGN detected in both the mid-infrared and X-ray domains \citep{maitra2019}. The analysis of further follow-up spectroscopic observations is ongoing. \cite{pennock2022} searched instead for dust-obscured AGN candidates through the application of a dimensionality-reduction algorithm to data from the visual to the radio domain. They found 14 whose extragalactic nature is subsequently confirmed with spectroscopic observations. Furthermore, an analysis of the properties of individual AGNs suggests that variability likely originates from the torus rather than the disc (Anih et al., in preparation).

\subsection{Stellar clusters}

The dense stellar population of 47 Tuc is superimposed on the distribution of a significantly lower number of field SMC stars. \cite{li2014} studied the type and distribution of stars within 47 Tuc concluding that the differences between the colours of sub-giant-branch and RGB stars in the inner and outer regions of the cluster confirm the presence of a second generation of stars. \cite{zhang2015}, by focusing instead on main-sequence stars, studied the luminosity and mass-functions using VMC data in the external regions of the cluster and data from the {\it Hubble} Space Telescope (HST) in the inner regions. They derived that mass segregation and tidal stripping are important for both populations. 

The deep and wide-area coverage of the Clouds offered by the VMC survey, allows us also to characterise and search for candidate stellar clusters. The procedure to identify and derive parameters like age, metallicity, extinction, distance and size necessitates the removal of field stars. This is performed using adjacent regions to the clusters whereas isochrones are used to extract the cluster parameters. In the LMC, \cite{piatti2014} found 65 stellar clusters with ages log($t$/yr) between 7.3 and 9.55. They are located within two VMC tiles, one south of the bar and the other at the South Ecliptic Pole. Furthermore, within the VMC tile containing 30 Doradus and a tile centred on the bar, \cite{piatti2015b} characterised more than 300 stellar clusters. In the SMC, five VMC tiles on the eastern side of the galaxy are studied resulting in a sample of 36 stellar clusters \citep{piatti2015a} of typically 12 pc in diameter. The oldest ones (log($t$/yr)$\ge$9.6) are preferentially located in the outer regions with the youngest clusters (log($t$/yr)$\sim$7.3) tracing the Bridge and Wing components of the galaxy. An excess of stellar clusters with ages log($t$/yr)<9 could be due to an episode of cluster formation triggered by a dynamical interaction with the LMC. In addition, 38 new stellar clusters are found within the south-west region of the SMC bar-like feature \citep{piatti2016}. These stellar clusters are small (3--7 pc in radius) and the VMC data are instrumental for recognising them if $E(B-V)$>0.6 mag.

Young star clusters still embedded in their parental nebulae and molecular clouds were, in the context of the VMC survey, first studied by \cite{romita2016}. Nearly 50 embedded cluster candidates were discovered within the VMC tile containing 30 Doradus, which are generally more luminous and massive than those in the Milky Way. An automatic search for star clusters that are semi-resolved in the VMC deep images shows that the same tile contains about 700 cluster candidates. This procedure first removes stars from the images and then identifies the clusters from isophotes where integrated properties allow to filter out most of the contaminating sources (see \citealp{miller2024} for details). An analysis of the parameters of the clusters (size and brightness) is ongoing as well as a dedicated study of multiple stellar clusters.

\section{Conclusions}
\label{conclusions}

This work presents the final data release of the VMC survey which acquired near-infrared imaging data on the Clouds for nine years. Additional programmes complement the VMC observations which together provide a high quality catalogue of about 64 million objects. The depth, sensitivity and wide-area of the VMC survey have allowed to study a variety of scientific topics pertaining to the history of formation and evolution of the Clouds, their 3D structure and kinematics. Many of the VMC sources will soon have spectra from the 4-metre Multi-Object Spectroscopic Telescope (4MOST; \citealp{dejong2019}, \citealp{cioni2019}) which has replaced VIRCAM on VISTA, and the Multi Object Optical and Near-infrared Spectrograph (MOONS; \citealp{cirasuolo2020}, \citealp{gonzalez2020}) at the Very Large Telescope, introducing additional dimensions in the study of stellar populations of the Clouds and their evolution. 

In the near-infrared, the {\it Euclid} space mission \citep{mellier2024} is producing images in the $Y$, $J$ and $H$ bands with a spatial resolution of 0.3 arcsec and a detection limit of 5$\sigma$ for point sources of 24 mag. The field-of-view is of 0.53 deg$^2$ and similarly to VISTA, Euclid is equipped with an array of 16 detectors. The Euclid observing programme includes the SMC while the LMC may be observed later in the mission following a call for proposals to optimise the observing strategy. Euclid observations will allow us to disentangle sources in the densest regions of the Clouds and to detect fainter stars in their outskirts. The advent of the {\it Roman} Space Telescope\footnote{https://science.nasa.gov/mission/roman-space-telescope/} will revolutionise the study of near-infrared stellar populations in the Clouds, as it is expected to provide deep images with a spatial resolution of $<0.2$ arcsec across wide sky areas. Contemporary with these projects, we will also be able to use data from subsequent data releases of the {\it Gaia} mission, providing in particular quantities (magnitudes and proper motions) for sources in the inner regions of the Clouds \citep{weingrill2023}, and from the Vera C. Rubin Observatory Legacy Survey of Space and Time, which will map stellar populations of the Clouds in multiple optical filters and with a high cadence \citep{street2023}. This incredible data set will allow us to make significant progress in understanding the nature of stellar populations and the history of our neighbouring galaxies.

\begin{acknowledgements}

This project is based on observations collected at the European Organisation for Astronomical Research in the Southern Hemisphere under ESO programme(s) 179.B-2003, 099.C-0773, 099.D-0194, 0100.C-0248, 0103.B-0783, 0103.D-0161, 105.2042, 105.2043, 106.2107, 108.222A, 108.223E, 109.230A, 109.231H, and 110.259F. We thank all observers and support staff at ESO contributing to the realisation of the programme. We thank the Cambridge Astronomy Survey Unit (CASU) and the Wide Field Astronomy Unit (WFAU) in Edinburgh for providing calibrated data products under the support of the Science and Technology Facility Council (STFC) in the UK. In particular, we thank J. Lewis for developing the VDFS, C. Gonzalez Fernandez, E. Gonzales Solares and A. Kupcu Yoldas for processing the images and answer various queries.

\end{acknowledgements}

\bibliographystyle{aa}
\bibliography{my}

\begin{appendix}

\section{DR7 footprint}
\label{maps}
\begin{figure}
\centering
\includegraphics[width=\hsize]{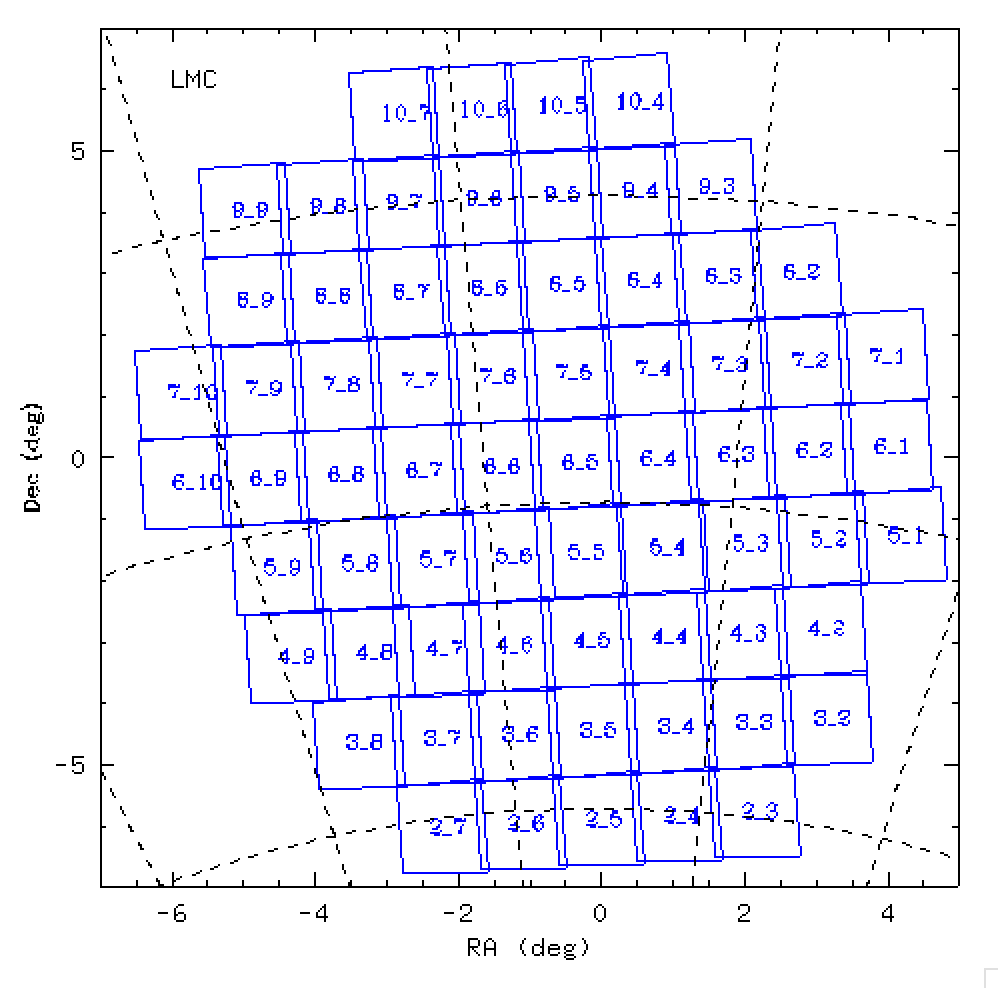}
\caption{Distribution of VISTA tiles across the LMC area. Tile names refer to LMC row\_column. The map is centred at RA=80.30 deg and Dec= --72.42 deg. North is at the top and East to the left.}
\label{lmcmap}
\end{figure}

\begin{figure}
\centering
\includegraphics[width=\hsize]{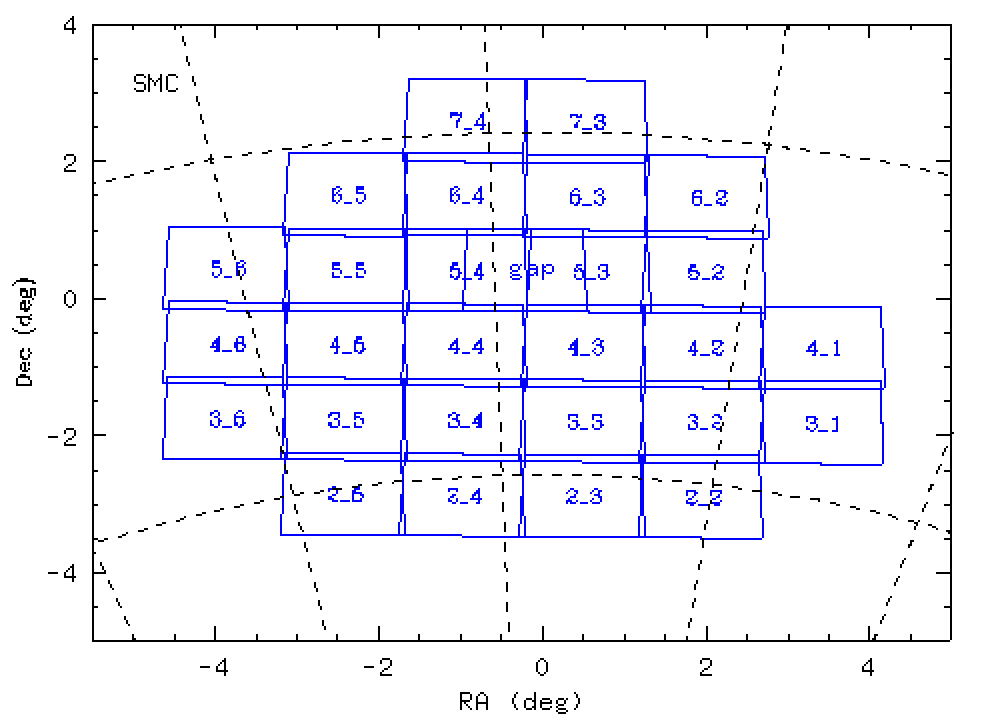}
\caption{Distribution of VISTA tiles across the SMC area. Tile names refer to SMC row\_column. The map is centred at RA=13.05 deg and Dec= --69.27 deg. North is at the top and East to the left. Note the location of the SMC\_gap tile between tiles SMC 5\_3 and 5\_4.}
\label{smcmap}
\end{figure}

\begin{figure}
\centering
\includegraphics[width=\hsize]{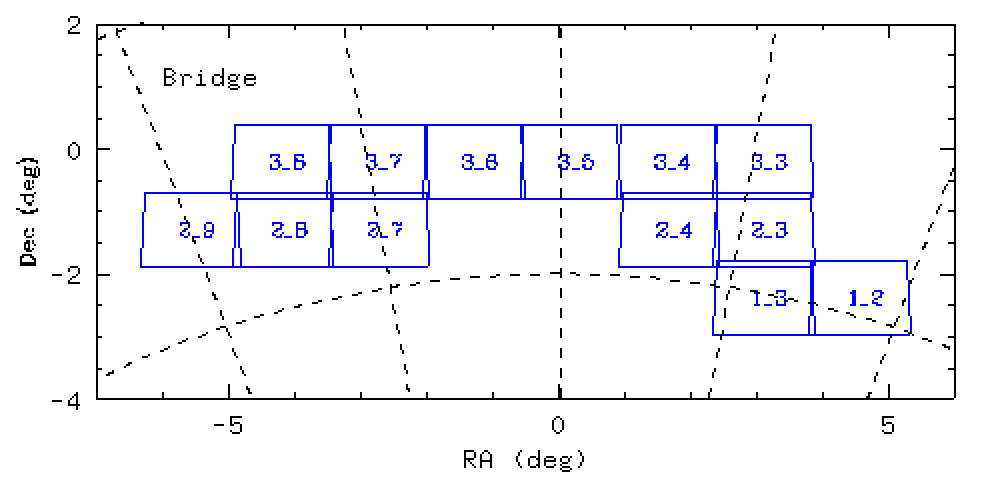}
\caption{Distribution of VISTA tiles across the Bridge area. Tile names refer to BRI row\_column. The map is centred at RA=45 deg and Dec= --73 deg. North is at the top and East to the left.}
\label{bridgemap}
\end{figure}

The distributions of VISTA tiles across the LMC, SMC and Bridge areas covered in DR7 are shown in Figs.\,\ref{lmcmap}, \ref{smcmap} and \ref{bridgemap}, respectively. In addition there are two tiles on the Stream located respectively above the SMC (tile STR 2\_1) centred at (00:11:59.424, --64:39:30.960) and above the Bridge (tile STR 1\_1) centred at (03:30:03.936, --64:25:23.880). The exact coordinates of all tile centres are given in \cite{cioni2011}, except for tile LMC 7\_1 which was not originally planned, this tile is centred at (04:40:09.167, --67:18:19.800), and for the SMC-gap tile from the additional programmes, see Sect.\,\ref{gapsection}.

\section{Low-quality images}

Table \ref{lowquality1} provides information about the VISTA tile images that were obtained outside the VMC sky-quality criteria, including those with significant (0.1 mag or larger) zero-point variations across the tile. Each line refers to a single tile observation and indicates: the tile, the date of observation, the filter, and the type (deep or shallow), which refers to the exposure time of the individual pawprints. Then it shows the FWHM, ellipticity, zero point and airmass, as well as the reason(s) of the low-quality, namely which attribute(s) were violated with respect to the requirements e.g. FWHM, ellipticity, transparency or airmass. The wordings ``zero point" refers to observations for which the zero-point variation among the six pawprints that make up a tile is $\geq$0.1 mag, whereas ``exposure time" indicates that observations were acquired with DIT=1 s rather than 5 s as for all others tiles in the $K_\mathrm{s}$ band. An asterisk next to a value of FWHM or ellipticity indicates that the value corresponding to the worst pawprint is listed instead of the one provided by CASU for the combined tile. This is because the latter (without a dispersion associated to it) may not reflect the poor quality of the data. For observations of tiles with a pointing different from that of the other observations of the same tile a footnote is linked to the entry for the low-quality reason. Table \ref{lowquality2} reports the same information as in Table \ref{lowquality1} for the additional programmes where the corresponding ESO programme identification is also listed.

Table \ref{pawprints} reports the same quantities for single pawprints, from both the VMC and the additional programmes, that could not result into complete tiles. The format of Table \ref{pawprints} is the same as that of Table \ref{lowquality2}. Among the low-quality reasons the wording ``<5 images" indicates pawprints with an incomplete number of dithering offsets.

\begin{table*}
	\small
	\setlength{\tabcolsep}{2.5pt}
	\caption{Low-quality images from the VMC programme.}
	\label{lowquality1}
	\centering
\begin{tabular}{lcclccccl}
	\hline\hline
	Tile & Date & Filter & Type & FWHM & Ellipticity & Zero Point ($\sigma$) & Airmass & Reason \\
	 & & & & ($^{\prime\prime}$) & & (mag) & & \\
	\hline
	LMC 2\_3 & 2016-08-20 & $J$ & deep & 1.51 & 0.06 & 23.74 (0.01) & 1.630 & FWHM \\
	 & 2017-11-20 & $K_\mathrm{s}$ & deep & 1.36 & 0.04 & 22.93 (0.01) & 1.577 & FWHM \\
	 & 2016-11-17 & $K_\mathrm{s}$ & deep & 1.34 & 0.05 & 22.71 (0.10) & 1.650 & zero point \\
	 & 2017-10-04 & $K_\mathrm{s}$ & deep & 1.04 & 0.04 & 22.73 (0.11) & 1.682 & zero point \\
	LMC 2\_4 & 2015-11-09 & $Y$ & deep & 1.58 & 0.07 & 23.48 (0.01) & 1.572 & FWHM \\
	 & 2013-10-15 & $K_\mathrm{s}$ & deep & 0.96 & 0.05 & 22.78 (0.13) & 1.617 & zero point \\
\hline
	\end{tabular}	
\end{table*}

\addtocounter{table}{-1}
\begin{table*}
	\small
	\setlength{\tabcolsep}{2.5pt}
	\caption{continued.}
	\centering
	\begin{tabular}{lcclccccl}
	\hline\hline
	Tile & Date & Filter & Type & FWHM & Ellipticity & Zero Point ($\sigma$) & Airmass & Reason \\
	 & & & & ($^{\prime\prime}$) & & (mag) & & \\
	\hline
	 & 2012-01-24 & $K_\mathrm{s}$ & deep & 0.98 & 0.23$^*$ & 23.08 (0.01) & 1.486 & ellipticity \\
	 & 2012-11-28 & $K_\mathrm{s}$ & deep & 1.31 & 0.04 & 23.06 (0.02) & 1.521 & FWHM \\
	LMC 4\_4 & 2014-12-03 & $Y$ & shallow & 1.31 & 0.05 & 23.50 (0.01) & 1.571 & FWHM \\
	 & 2014-12-03 & $J$ & shallow & 1.36 & 0.05 & 23.73 (0.01) & 1.529 & FWHM \\
	 & 2015-10-31 & $K_\mathrm{s}$ & deep & 1.80 & 0.08 & 23.04 (0.01) & 1.490 & FWHM \\
	LMC 4\_5 & 2013-10-14 & $Y$ & shallow & 0.82 & 0.06 & 23.62 (0.14) & 1.487 & zero point \\
\hline
	\end{tabular}
\end{table*}

\addtocounter{table}{-1}
\begin{table*}
	\small
	\setlength{\tabcolsep}{2.5pt}
	\caption{continued.}
	\centering
	\begin{tabular}{lcclccccl}
	\hline\hline
	Tile & Date & Filter & Type & FWHM & Ellipticity & Zero Point ($\sigma$) & Airmass & Reason \\
	 & & & & ($^{\prime\prime}$) & & (mag) & & \\
	\hline
	 & 2014-01-13 & $K_\mathrm{s}$ & deep & 1.00 & 0.06 & 23.00 (0.01) & 1.468 & FWHM \\
	LMC 5\_8 & 2017-01-26 & $Y$ & deep & 1.23 & 0.05 & 23.22 (0.11) & 1.497 & FWHM, zero point \\
	 & 2018-01-24 & $K_\mathrm{s}$ & deep & 0.78 & 0.04 & 22.74 (0.10) & 1.444 & zero point \\
	LMC 5\_9 & 2015-11-09 & $Y$ & deep & 2.10 & 0.06 & 23.48 (0.01) & 1.439 & FWHM \\
	 & 2016-02-17 & $Y$ & deep & 1.53 & 0.06 & 23.18 (0.01) & 1.486 & FWHM \\
	 & 2016-03-23 & $Y$ & shallow & 1.42 & 0.08 & 23.45 (0.01) & 1.539 & FWHM \\
\hline
	\end{tabular}
\end{table*}
	
\addtocounter{table}{-1}
	\begin{table*}
	\small
	\setlength{\tabcolsep}{2.5pt}	
	\caption{continued.}
	\centering
	\begin{tabular}{lcclccccl}
	\hline\hline
	Tile & Date & Filter & Type & FWHM & Ellipticity & Zero Point ($\sigma$) & Airmass & Reason \\
	 & & & & ($^{\prime\prime}$) & & (mag) & & \\
	\hline	
	LMC 7\_9 & 2017-11-12 & $Y$ & deep & 1.98 & 0.05 & 23.20 (0.09) & 1.370 & FWHM \\
	 & 2017-11-18 & $Y$ & deep & 1.78 & 0.08 & 23.32 (0.01) & 1.371 & FWHM \\
	 & 2018-01-25 & $Y$ & shallow & 1.09 & 0.06 & 22.66 (0.40) & 1.384 & zero point \\
	 & 2018-01-11 & $J$ & shallow & 0.67 & 0.07 & 23.33 (0.66) & 1.571 & zero point \\
	 & 2014-02-06 & $K_\mathrm{s}$ & deep & 0.84 & 0.05 & 22.46 (0.15) & 1.524 & zero point \\
	 & 2018-02-18 & $K_\mathrm{s}$ & deep & 1.29 & 0.10 & 22.87 (0.01) & 1.426 & FWHM \\
\hline
	\end{tabular}
\end{table*}	
	
\addtocounter{table}{-1}
	\begin{table*}
	\small
	\setlength{\tabcolsep}{2.5pt}	
	\caption{continued.}
	\centering
	\begin{tabular}{lcclccccl}
	\hline\hline
	Tile & Date & Filter & Type & FWHM & Ellipticity & Zero Point ($\sigma$) & Airmass & Reason \\
	 & & & & ($^{\prime\prime}$) & & (mag) & & \\
	\hline		
	 & 2011-02-18 & $J$ & deep & 1.30 & 0.05 & 23.43 (0.01) & 1.403 & FWHM \\
	 & 2011-03-11 & $J$ & shallow & 1.42 & 0.05 & 23.42 (0.01) & 1.401 & FWHM \\
	 & 2011-11-11 & $J$ & shallow & 1.98 & 0.06 & 23.76 (0.04) & 1.321 & FWHM \\
	 & 2012-09-18 & $J $& shallow & 1.23 & 0.06 & 23.80 (0.01) & 1.384 & FWHM \\
	 & 2011-03-01 & $K_\mathrm{s}$ & deep & 1.10 & 0.05 & 22.69 (0.31) & 1.384 & zero point \\
	 & 2013-01-06 & $K_\mathrm{s}$ & deep & 1.46 & 0.04 & 23.07 (0.01) & 1.356 & FWHM \\
\hline
	\end{tabular}
	\end{table*}

\addtocounter{table}{-1}
\begin{table*}
	\small
	\setlength{\tabcolsep}{2.5pt}
	\caption{continued.}
	\centering
	\begin{tabular}{lcclccccl}
	\hline\hline
	Tile & Date & Filter & Type & FWHM & Ellipticity & Zero Point ($\sigma$) & Airmass & Reason \\
	 & & & & ($^{\prime\prime}$) & & (mag) & & \\
	\hline
          & 2013-08-22 & $K_\mathrm{s}$ & deep & 1.21 & 0.05 & 23.04 (0.01) & 1.604 & FWHM \\
          & 2014-10-20 & $K_\mathrm{s}$ & shallow & 0.93 & 0.05 & 22.74 (0.10) & 1.518 & zero point \\
          & 2014-11-26 & $K_\mathrm{s}$ & deep & 1.45 & 0.06 & 23.00 (0.01) & 1.545 & FWHM \\
        BRI 3\_3 & 2013-01-07 & $Y$ & deep & 1.35 & 0.06 & 23.48 (0.01) & 1.601 & FWHM \\
         & 2013-08-09 & $Y$ & deep & 1.32 & 0.05 & 23.47 (0.01) & 1.511 & FWHM \\
         & 2014-08-14 & $K_\mathrm{s}$ & deep & 1.23$^*$ & 0.05 & 23.02 (0.01) & 1.578 & FWHM \\
\hline
	\end{tabular}
\end{table*}
	
\addtocounter{table}{-1}
\begin{table*}
	\small
	\setlength{\tabcolsep}{2.5pt}
	\caption{continued.}
	\centering
	\begin{tabular}{lcclccccl}
	\hline\hline
	Tile & Date & Filter & Type & FWHM & Ellipticity & Zero Point ($\sigma$) & Airmass & Reason \\
	 & & & & ($^{\prime\prime}$) & & (mag) & & \\
	\hline
	 & 2013-06-29 & $K_\mathrm{s}$ & deep & 1.07 & 0.05 & 23.05 (0.01) & 1.675 & FWHM \\
	 & 2014-07-10 & $K_\mathrm{s}$ & deep & 1.68 & 0.06 & 23.05 (0.01) & 1.670 & FWHM \\
	SMC 4\_1 & 2013-07-29 & $Y$ & shallow & 1.34$^*$ & 0.07 & 23.46 (0.01) & 1.514 & FWHM \\
	 & 2013-07-18 & $J$ & deep & 0.78 & 0.07 & 23.84 (0.20) & 1.642 & zero point \\
	 & 2013-07-29 & $J$ & shallow & 1.33 & 0.07 & 23.70 (0.01) & 1.502 & FWHM \\
	 & 2013-07-23 & $K_\mathrm{s}$ & shallow & 1.11 & 0.05 & 23.06 (0.01) & 1.498 & FWHM \\
\hline
	\end{tabular}	
\end{table*}

\addtocounter{table}{-1}
\begin{table*}
	\small
	\setlength{\tabcolsep}{2.5pt}
	\caption{continued.}
	\centering
	\begin{tabular}{lcclccccl}
	\hline\hline
	Tile & Date & Filter & Type & FWHM & Ellipticity & Zero Point ($\sigma$) & Airmass & Reason \\
	 & & & & ($^{\prime\prime}$) & & (mag) & & \\
	\hline
	 & 2013-10-07 & $K_\mathrm{s}$ & deep & 1.04 & 0.09 & 23.05 (0.01) & 1.519 & FWHM \\
	SMC 5\_6 & 2011-09-14 & $Y$ & shallow & 1.15 & 0.06 & 23.54 (0.01) & 1.589 & FWHM \\
	 & 2011-12-07 & $Y$ & deep & 1.46 & 0.08 & 23.55 90.01) & 1.570 & FWHM \\
	 & 2011-12-21 & $Y$ & deep & 1.59 & 0.06 & 23.49 (0.01) & 1.514 & FWHM \\
	 & 2011-12-22 & $Y$ & deep & 1.15 & 0.06 & 23.48 (0.01) & 1.534 & FWHM \\
	 & 2011-12-24 & $Y$ & deep & 0.95 & 0.07 & 22.62 (0.47) & 1.512 & zero point \\
\hline
	\end{tabular}	
	\tablefoot{$^{*}$ value from the worst pawprint in the tile; $^{**}$ central tile coordinates differ from other epoch.}
\end{table*}

\begin{table*}
	\small
	\setlength{\tabcolsep}{2.5pt}
	\caption{Low-quality images of VMC tiles from the additional programmes.}
	\label{lowquality2}
	\centering
	\begin{tabular}{lcclccccll}
	\hline\hline
	Tile & Date & Filter & Type & FWHM & Ellipticity & Zero Point ($\sigma$) & Airmass & Reason & Programme\\
	 & & & & ($^{\prime\prime}$) & & (mag) & & \\
	\hline
	LMC 3\_4 & 2023-01-13 & $K_\mathrm{s}$ & deep & 1.23 & 0.05 & 23.03 (0.01) & 1.525 & FWHM & 110.259F \\
	LMC 4\_4 & 2021-12-16 & $Y$ & deep & 1.17 & 0.07 & 23.03 (0.31) & 1.504 & zero point & 108.223E \\
	LMC 5\_1 & 2023-01-14 & $K_\mathrm{s}$ & deep & 1.22 & 0.06 & 23.03 (0.01) & 1.436 & FWHM & 110.259F \\
	LMC 5\_2 & 2021-12-01 & $K_\mathrm{s}$ & deep & 0.99 & 0.07 & 23.05 (0.01) & 1.549 & FWHM & 108.222A \\
	LMC 5\_3 & 2021-12-05 & $K_\mathrm{s}$ & deep & 0.88 & 0.06 & 21.78 (0.44) & 1.535 & FWHM, zero-point & 108.222A \\
	LMC 5\_6 & 2021-01-29 & $K_\mathrm{s}$ & deep & 1.24 & 0.07 & 23.05 (0.01) & 1.562 & FWHM & 106.2107 \\
\hline
	\end{tabular}	
	\tablefoot{$^{*}$ value from the worst pawprint in the tile; $^{**}$ central tile coordinates differ from other epochs.}	
\end{table*}

\begin{table*}
	\small
	\setlength{\tabcolsep}{3pt}
	\caption{Quality of single pawprint images of VMC tiles and tiles from the additional programmes.}
	\label{pawprints}
	\centering
	\begin{tabular}{lcclccccll}
	\hline\hline
	Pawprint & Date & Filter & Type & FWHM & Ellipticity & Zero Point ($\sigma$) & Airmass & Reason & Programme\\
	 & & & & ($^{\prime\prime}$) & & (mag) & & \\
	 \hline
	LMC 2\_6 & 2018-02-26 & $K_\mathrm{s}$ & deep & 0.92 & 0.09 & 22.87 (0.01) & 1.603 & & 179.B-2003 \\
	 & 2018-02-26 & $K_\mathrm{s}$ & deep & 0.93 & 0.07 & 22.88 (0.01) & 1.617 & & 179.B-2003 \\
	LMC 3\_3 & 2016-11-17 & $K_\mathrm{s}$ & deep & 1.31 & 0.12 & 22.96 (0.01) & 1.534 & FWHM & 179.B-2003 \\
	 & 2016-11-17 & $K_\mathrm{s}$ & deep & 1.52 & 0.09 & 23.02 (0.02) & 1.535 & FWHM, <5 images & 179.B-2003 \\
	 LMC 3\_5 & 2010-12-05 & $K_\mathrm{s}$ & deep & 0.92 & 0.07 & 22.97 (0.02) & 1.577 & <5 images & 179.B-2003 \\
	  & 2011-09-01 & $K_\mathrm{s}$ & deep & 0.77 & 0.08 & 22.92 (0.01) & 1.655 & <5 images & 179.B-2003 \\
\hline
	\end{tabular}
\end{table*}

\addtocounter{table}{-1}
\begin{table*}
	\small
	\setlength{\tabcolsep}{3pt}
	\caption{continued.}
	\centering
	\begin{tabular}{lcclccccll}
	\hline\hline
	Pawprint & Date & Filter & Type & FWHM & Ellipticity & Zero Point ($\sigma$) & Airmass & Reason & Programme\\
	 & & & & ($^{\prime\prime}$) & & (mag) & & \\
	 \hline
	   & 2014-03-03 & $Y$ & shallow & 1.16 & 0.06 & 23.47 (0.03) & 1.546 & humidity & 179.B-2003 \\
	  LMC 7\_3 & 2011-02-19 & $K_\mathrm{s}$ & deep & 0.84 & 0.05 & 22.86 (0.01) & 1.605 & transparency & 179.B-2003 \\
	   & 2011-02-19 & $K_\mathrm{s}$ & deep & 0.94 & 0.04 & 22.67 (0.02) & 1.629 & transparency & 179.B-2003 \\
	   & 2011-02-19 & $K_\mathrm{s}$ & deep & 0.99 & 0.04 & 22.66 (0.02) & 1.661 & transparency & 179.B-2003 \\	   
	   & 2011-02-19 & $K_\mathrm{s}$ & deep & 0.94 & 0.06 & 22.17 (0.20) & 1.689 & zero-point & 179.B-2003 \\
	   & 2011-02-19 & $K_\mathrm{s}$ & deep & 1.11 & 0.05 & 20.51 (0.73) & 1.724 & zero-point, <5 images, airmass & 179.B-2003 \\
\hline
	\end{tabular}
\end{table*}	

\addtocounter{table}{-1}
\begin{table*}
	\small
	\setlength{\tabcolsep}{3pt}
	\caption{continued.}
	\centering
	\begin{tabular}{lcclccccll}
	\hline\hline
	Pawprint & Date & Filter & Type & FWHM & Ellipticity & Zero Point ($\sigma$) & Airmass & Reason & Programme\\
	 & & & & ($^{\prime\prime}$) & & (mag) & & \\
	 \hline
	  & 2011-11-11 & $K_\mathrm{s}$ & shallow & 1.83 & 0.07 & 23.04 (0.01) & 1.344 & FWHM & 179.B-2003 \\
	  & 2011-11-11 & $K_\mathrm{s}$ & shallow & 1.82 & 0.08 & 23.05 (0.02) & 1.345 & FWHM & 179.B-2003 \\
	  & 2011-11-11 & $K_\mathrm{s}$ & shallow & 1.59 & 0.08 & 23.06 (0.01) & 1.352 & FWHM & 179.B-2003 \\
	LMC 9\_8 & 2018-01-11 & $K_\mathrm{s}$ & shallow & 0.85 & 0.06 & 22.84 (0.01) & 1.308 & transparency & 179.B-2003 \\
	 & 2018-01-11 & $K_\mathrm{s}$ & shallow & 0.77 & 0.06 & 22.73 (0.03) & 1.308 & transparency & 179.B-2003 \\
	 & 2018-01-11 & $K_\mathrm{s}$ & shallow & 0.83 & 0.05 & 22.36 (0.07) & 1.312 & transparency & 179.B-2003 \\
\hline
	\end{tabular}
\end{table*}	

\addtocounter{table}{-1}
\begin{table*}
	\small
	\setlength{\tabcolsep}{3pt}
	\caption{continued.}
	\centering
	\begin{tabular}{lcclccccll}
	\hline\hline
	Pawprint & Date & Filter & Type & FWHM & Ellipticity & Zero Point ($\sigma$) & Airmass & Reason & Programme\\
	 & & & & ($^{\prime\prime}$) & & (mag) & & \\
	 \hline
	SMC 3\_1 & 2012-09-08 & $K_\mathrm{s}$ & shallow & 1.16 & 0.08 & 23.01 (0.02) & 1.635 & FWHM & 179.B-2003 \\
	 & 2012-09-08 & $K_\mathrm{s}$ & shallow & 1.12 & 0.09 & 23.04 (0.03) & 1.625 & FWHM, <5 images & 179.B-2003 \\
	SMC 3\_3 & 2010-12-14 & $K_\mathrm{s}$ & shallow & 1.05 & 0.06 & 22.97 (0.02)  & 1.669 & FWHM & 179.B-2003 \\
	 & 2010-12-14 & $K_\mathrm{s}$ & shallow & 0.99 & 0.06 & 22.99 (0.02) & 1.676 & FWHM, <5 images & 179.B-2003 \\
	SMC 3\_4 & 2014-08-15 & $J$ & shallow & 0.87 & 0.06 & 23.68 (0.01) & 1.562 & & 179.B-2003 \\
	 & 2014-08-15 & $J$ & shallow & 0.84 & 0.05 & 23.68 (0.01) & 1.555 & & 179.B-2003 \\
\hline
	\end{tabular}
\end{table*}

\end{appendix}

\end{document}